\documentclass[prl,twocolumn,,aps,showpacs]{revtex4} 

\usepackage{graphicx}
\usepackage{dcolumn}
\usepackage{bm}
\usepackage{amssymb}
\usepackage{hyperref}

\begin{document}

\title{Economic resilience from input--output susceptibility improves predictions of economic growth and recovery}

\author{Peter Klimek$^{1,2}$, Sebastian Poledna$^{2,3}$, Stefan Thurner$^{1,2,3,4,*}$}
\affiliation{$^1$Section for Science of Complex Systems, CeMSIIS, Medical University of Vienna, Spitalgasse 23, A-1090, Austria\\
$^2$Complexity Science Hub Vienna, Josefst\"adter Strasse 39, A-1080 Vienna, Austria\\
$^3$IIASA, Schlossplatz 1, A-2361 Laxenburg, Austria\\
$^4$Santa Fe Institute, 1399 Hyde Park Road, Santa Fe, NM 85701, USA\\
$^*$To whom correspondence should be addressed. E-mail: stefan.thurner@muv.ac.at
}

\begin{abstract}
Modern macroeconomic theories were unable to foresee the last Great Recession and could neither predict its prolonged duration nor the recovery rate. They are based on supply-demand equilibria that do not exist during recessionary shocks. Here we focus on resilience as a nonequilibrium property of networked production systems and develop a linear response theory for input-output economics. By calibrating the framework to data from 56 industrial sectors in 43 countries between 2000 and 2014, we find that the susceptibility of individual industrial sectors to economic shocks varies greatly across countries, sectors, and time. We show that susceptibility-based predictions that take sector- and country-specific recovery into account, outperform---by far---standard econometric growth-models. Our results are analytically rigorous, empirically testable, and flexible enough to address policy-relevant scenarios. We illustrate the latter by estimating the impact of recently imposed tariffs on US imports (steel and aluminum) on specific sectors across European countries.
\end{abstract}

\maketitle

In 2008 several advanced economies were hit by the largest recessionary shock in history \cite{IMF09}.
This Great Recession was followed by an even more remarkable event, namely a puzzlingly slow rate of economic recovery \cite{Barnett14}.
Economists not only got the likelihood of a crisis of this severity wrong, as Paul Krugman famously noted, but also how fast we would recover from it \cite{Bryson14}.
Efforts to understand the origin of this blind spot in economic theory and its failure to predict systemic events have fueled the interest in how economic systems absorb shocks and how they recover \cite{Schweitzer09, Garas10, Haldane11, Farmer12}.
Our lack of understanding of economic resilience \cite{Martin11} has been explained by a fundamental mismatch between macroeconomic theories and the reality of how markets work, especially in the presence of extreme events \cite{Kirman10}.
General equilibrium theory holds that economic growth is characterized by a balance of demand and supply which results in prices that signal an overall equilibrium \cite{Arrow54, Smets07, Mitra08}.
However, the crisis was  a ``story of contagion, interdependence, interaction, networks, and trust'' \cite{Kirman10} that led these equilibrium assumptions {\it ad absurdum}.
The inappropriate use of equilibrium concepts in economics 
in the context of extreme events was pointed out some time ago \cite{Farmer05}.
So far, the {\em only} way to address and study economic non-equilibrium 
are highly stylized statistical models of money exchange \cite{Gallegatti06} and computer simulations \cite{Farmer09,Klimek14}.

In physics, non-equilibrium systems are equally hard to understand and control. 
Aside from some seminal contributions \cite{Onsager31, Kubo66, Nicolis77, Hanel14}, a unified framework for out-of-equilibrium phenomena has yet to be found \cite{Jarzynski15}.
However, to understand how systems in equilibrium behave in response to shocks has been successfully addressed within the framework of linear response theory (LRT). 
According to LRT, an external force, $X(t)$, acting on a system induces a proportional flux, $J(t)=\rho X(t)$.
The proportionality is given by transport coefficients or susceptibilities, $\rho$, 
that are formally related to the decay of the system's equilibrium autocorrelation functions---the so-called Green--Kubo relations \cite{Green54, Kubo57}.
LRT provides the theoretical basis for many linear phenomenological laws that constitute the core of each high school physics curriculum, such as Ohm's law, Newtonian viscosity, or magnetic and electric susceptibilities, see Supplementary Material (SM) Note S1 and Table S1 \ref{LRTLaws}.

In the following we give an intuitive account of LRT.
Imagine someone hands you a serving tray with an elaborate house of cards on it.
Which card will fall first and cause the collapse of the house?
You try to answer this question by doing minimal damage: you slightly nudge the tray and observe how the cards respond.
If a tiny nudge moves certain cards, you might conclude that those are the first to fall if the tray was pushed harder.
The first cards that you observe to move you call the most ``susceptible'' to the shock (nudge).
If you observe no movements of cards whatsoever, you might be tempted to apply a stronger kick; the cards could be glued together.
A similar way of reasoning underlies the theory of linear response.
In the language of statistical mechanics, the tiny nudge that you initially apply plays the role of equilibrium fluctuations. 
These fluctuations may or may not move certain cards as a response---they induce a flux.
One then assumes that this response is proportional to the magnitude of the nudge, the proportionality being described by transport coefficient.
We will show that the same rationale can be used to study the response of Leontief IO economies \cite{Leontief86} to large shocks.
The resulting framework is applied to IO data from 56 industrial sectors in 43 countries between 2000 and 2014 \cite{Timmer15}; see SM Note S2.
We identify economic sectors with high susceptibility to production shocks in other parts of the economy. 
In the picture above, highly susceptible sectors correspond to cards that move first.
We show that the lack of recovery after the Great Recession can be related to the susceptibility of individual sectors.

How Leontief economies respond to shocks has been studied in a number of works briefly summarized as follows \cite{Long83, Haimes01, Contreras14, Acemoglu16, He17}.
Considering a (variant of a) Leontief IO economy, a shock is specified using varying degrees of external assumptions.
It is then studied how the economy relaxes to the old or new equilibrium configuration, unless yet another shock is assumed.
Our approach follows a completely different strategy.
We consider shocks that drive the economy away from its equilibrium into a non-equilibrium stationary state.
This stationary state is different from the original equilibrium state and the equilibrium state implied by the perturbed productivity or technology.
Instead, the new state is characterized by the system trying to achieve a balance between two opposing forces, namely (i) a relaxation to the (unaltered) equilibrium state and (ii) the direct and indirect influences of the external shock that actively drives the system away from equilibrium.
We call an economy in such a state a {\it driven} economy.
Market participants (sectors) in a driven IO economy incorporate the external shock in their production functions without altering their demand or required input from other sectors.
The perturbed output of these sectors then propagates along the IO network to other sectors, thereby driving the economy into a new non-equilibrium stationary state.

In this work we develop an analytic and empirically testable framework for the non-equilibrium response and recovery of severely disrupted economies. 
For the first time we formulate a theory of linear response for input--output (IO) economics \cite{Evans90, Leontief86}.
As in statistical physics, in the economic context LRT serves as a firm analytic link between the microscopic equilibrium fluctuations of a system and its macroscopic out-of-equilibrium response to large shocks.

The underlying ideas of LRT have been exploited in other contexts, such as the theory of linear time-invariant systems, 
with applications in signal processing and control theory \cite{Phillips13}.
In econometrics, impulse response functions describe how external shocks drive macroeconomic variables such as output, consumption, or employment in vector autoregressive (VAR) models \cite{Pesaran98, Lutkepohl08}.
Instead of studying the relaxation dynamics of macroeconomic variables within highly stylized VAR models,  
we focus on structural characteristics of dynamical IO matrices that capture the interactions between economic sectors.

\begin{figure}[t]
\begin{center}
 \includegraphics[width = 0.50\textwidth, keepaspectratio = true]{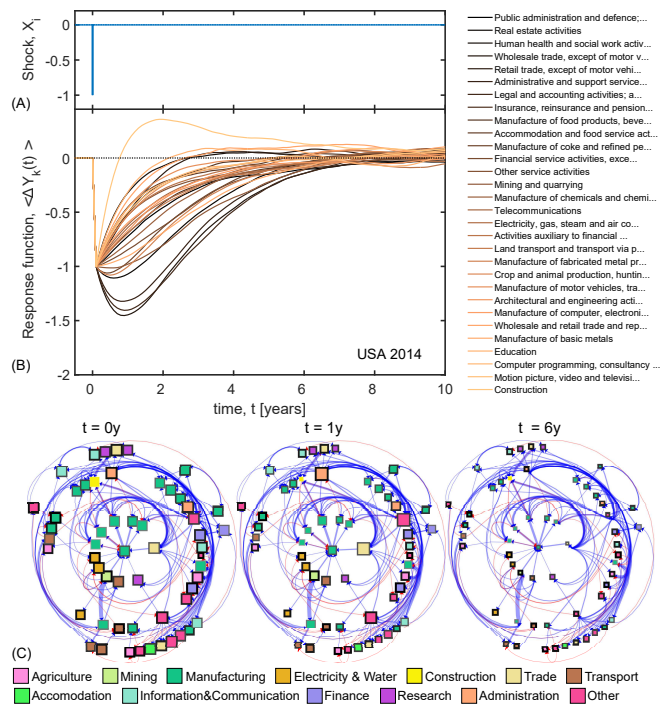}
\end{center}
 \caption{Visualization of response curves. 
 (A) An impulse shock of unit size is applied in year $t=2014$ to every sector, $i$, in the USA. 
 In response, the output of each sector is driven from its equilibrium value, given by $\langle \Delta Y_i(t')  \rangle_X= 0$. 
 (B) Every line corresponds to one of the 30 largest sectors, ordered according to their susceptibility to the shock 
(i.e. the area between the response curve and the dotted line that represents the equilibrium value). 
The sectors with the largest impact are public administration, real estate activities, human health, and wholesale trade. 
On the other end of the scale we find the construction sector, that after the initial shock profits from the disruptive event. 
Note the time scale. Depending on the sector, full economic recovery might take up to six to ten years.
(C) A network visualization of the backbone of the susceptibility matrix $\rho^c_{ij}(t)$ for the USA in 2014 is shown. 
Nodes are sectors and blue (red) weighted links indicate positive (negative) susceptibilities. 
Node colors show groups of sectors (see SM Table S2) and thickness of the node border gives the sum of the weights of the incoming links.
Node sizes are inversely proportional to the values of the response functions in (B) for $t'=t$, 
$t'=t+1$ and $t'=t+6$  years after the shock was applied. Source data are provided as a Source Data file.
}
 \label{response}
\end{figure}

\section*{Results}
   
   \subsection{Obtaining economic susceptibilities from input--output data}

Our formalism provides a quantitative and data-driven method to benchmark individual countries and 
production sectors in terms of their economic susceptibilities to shocks; see Methods.
To illustrate the method, we measure country and sector level economic susceptibilities respectively, 
by using the world input--output database (WIOD) \cite{Timmer15}.
We consider data for 56 sectors in 43 countries between 2000 and 2014.
For each country, $c$, and year, $t$, we extract demands $D_i(t)$, technical coefficients $A_{ij}(t)$, and outputs $Y_i(t)$, where subscripts refer to sectors.
Our aim is to compute the economic susceptibility matrix for a country and year, $\rho^c_{ij}(t)$.
Here $t$ denotes the year the data was taken to compute $\rho^c_{ij}(t)$.
Based on data from $t$, we model output changes forward in time on a scale denoted by $t'>t$.
We numerically integrate the stochastic differential equation for a Leontief economy, Eq. (\ref{SLIOM}), in the absence of an external shock ($X(t')=0$ for all $t' \geq t$). Now the time-lagged equilibrium correlation functions between two sectors in Eq. (\ref{GenResponse}) can be computed.
The entries in $\rho^c_{ij}(t)$ correspond to the area under the curve of these correlation functions when plotted as a function of the time lag in Eq. (\ref{GenResponse}).
Susceptibilities of individual sectors $\rho^c_{i}(t)$ are the column-wise sums of matrix $\rho^c_{ij}(t)$.

Response curves of individual sectors are obtained by integrating the correlation functions under specific shocks.
In Fig. \ref{response} we assume an impulse demand shock of unit size applied at time $t'=t$, $X_i(t') = \delta(t'-t)$ in each sector $i$ (A), 
leading to different response curves for each sector in the USA in 2014 (B).
The shock is applied at $t=2014$ and results in the same large decrease of output in all sectors immediately after $t$.
For $t'>t$, there appear substantial differences between sectors.
For some sectors the shock is amplified, such as for public administration, real estate activities, health, or wholesale trade.
Other sectors immediately start to rebound from the shock, for instance, the various manufacturing sectors.
The fastest rebound is observed for the construction sector, where production even exceeds the equilibrium level ($0$) for an extended period of time.
Overall, it can take up to six to ten years for each sector to return to its equilibrium state (sectoral recovery time).
Whether a shock is amplified or suppressed in a sector depends on the structure of the susceptibility matrix $\rho^c_{ij}(t)$, see Fig. \ref{response}(C).
There we show the backbone of $\rho^c_{ij}(t)$ (obtained after applying the disparity filter with $p=0.05$ \cite{Serrano09}) as a directed weighted network.
Blue (red) links show positive (negative) susceptibilities. Node colors indicate groups of similar sectors, thickness of the node border is proportional to the sum of the weights of all incoming links, see SM Table S2; 
node sizes are inversely proportional to the values of the response functions in Fig. \ref{response}(B) at a particular point in time.
We show three snapshots of this network at the time when the initial shock is applied ($t'=t$), and one ($t'=t+1$), and six ($t'=t+6$) years afterwards. 
Figure \ref{response}(C) shows that some but not all of the sectors with a particularly strong shock amplification tend to be among those with a large 
number of incoming links (and weigths thereof), compare for instance the administration (large shock, many incoming links with strong weights) and construction (almost negative shock amplification, small number of incoming links) sectors.  

We apply the  above procedure for every year $t$ (where the shock is applied), every country $c$, and every sector $i$, 
to compute a susceptibility value, $\rho^c_i(t)$ .
The {\em average country susceptibilities}, $\rho^c=\langle \rho^c_i(t) \rangle_i $, are obtained by averaging $\rho^c_i(t)$ over all sectors $i$ and years $t$, see SM Fig. S1. 
The higher the values of $\rho^c$, the higher is the chance that any sector $i$ in $c$ will be impacted by a shock in any other sector $j$.
We find similar levels of susceptibility in a large number countries across Europe, North America, and China.
Substantially smaller susceptibilities are found for Croatia, Greece, Malta, and Luxembourg.
For those countries, our findings suggest a higher production concentration in a smaller number of sectors and consequently a smaller exposure to cascading impacts between different sectors (within the country).
At the other end of the spectrum, it is striking to see that four out of the five BRICS countries appear as the most susceptible countries, namely Russia, China, India, and Brazil; data for South Africa is not included in the WIOD due to the lack of available data with sufficient quality \cite{Timmer16}.
This suggests that the sustained above-average growth of these countries in the last ten to twenty years
did not go along with the formation of resilient economic production structures.

In SM Fig. S2 we show the {\em output-weighted average sector susceptibility}, $\rho_i=\langle \rho^c_i(t) \rangle_c$, see also Table S2. 
Sectors with the highest susceptibilities include wholesale trade, administrative services, electricity, and financial service activities.
This means that if a country experiences an economic shock, those sectors are most likely to be affected by shocks in other sectors.
In contrast, we find that sectors like scientific research, activities of extraterritorial organizations, 
manufacture of transport equipment, or air and water transport are relatively immune to cascading events.

\subsection{Empirical validation of linear response theory for input--output economics}  

\begin{figure}[t]
\begin{center}
 \includegraphics[width = .45\textwidth, keepaspectratio = true]{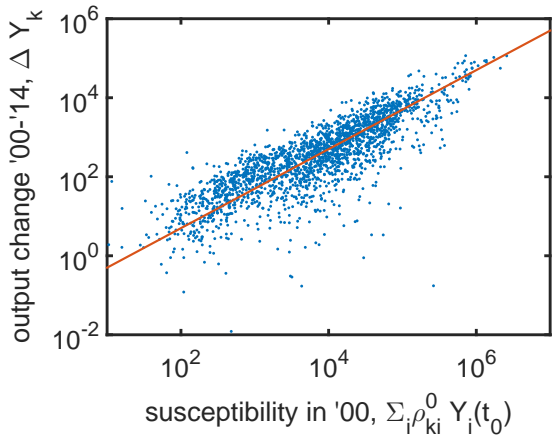}
\end{center}
 \caption{ Prediction of output fluctuations with economic susceptibility. 
 Under the assumption that each sector is driven by noise proportional to its output, we test the predictions that follow from the liner response framework,  Eq. (\ref{Regression}). We find good agreement between data and model ($r=0.83$); economic susceptibility is indeed predictive of future output fluctuations. The red line has a slope of one, indicating a linear relation. Source data are provided as a Source Data file.}
 \label{predict}
\end{figure}

We now show to what extent the economic susceptibilbricity matrix $\rho_{ij}$ is predictive of the size and direction of sectors' future output-changes.
First, it can be shown that the average size of sectoral output-changes can be predicted (out-of-sample) by means of sector-size-dependent random fluctuations.
To evaluate the linear response relation $\langle \Delta Y_k \rangle_X = \rho_{ki} X_i$ it is necessary to specify the shock, $X_i$.
A particularly simple assumption is that $X_i$ is itself 
noise with a magnitude proportional to the output of sector $i$, $X_i = \eta_i Y_i$, where $\eta_i$ has the same expectation value in each sector, 
$\langle \eta_i \rangle_i = \eta$.
The hypothesis is that if $\rho$ indeed captures structural characteristics of economies that relate to their recovery from shocks, 
one should be able to extract how violently $Y_k$ fluctuates in the future, based on its current susceptibility.
To test this, for every sector $k$ in every country $c$ we consider its 
annual absolute output change, $Y_k^c(t+1)-Y_k^c(t)$,  time-averaged over the range $t=2000,\dots,2013$, 
$\langle  \Delta Y_k^c  \rangle_t = (1/13)\sum_{t=2000}^{2013} \left( Y_k^c(t+1)-Y_k^c(t) \right)$. 
According to the above hypothesis, $\langle  \Delta Y_k^c  \rangle_t$ should be a function of susceptibility, $\rho^c_{ki}(t_0)$, and output, $Y_i(t_0)$, in the year $t_0=2000$.   
We therefore test the quality of the out-of-sample prediction given by 
\begin{equation}
\langle \Delta Y_k^c \rangle_t = \eta \sum_i \rho^c_{ki}(t_0) Y_i(t_0)\quad.
\label{Regression}
\end{equation}
Figure \ref{predict} shows that this relation indeed holds (Pearson's correlation coefficient of $r=0.83$).
This correlation is substantially stronger than the correlation between output change $\langle \Delta Y_k^c \rangle_t$ and output size $Y_k(t_0)$ alone ($r=0.56$).
Performing a linear regression of $\langle \Delta Y_k^c \rangle_t$ on $\sum_i \rho^c_{ki}(t_0) Y_i(t_0)$ {\em and} $Y_k(t_0)$ indeed yields a similar correlation as Eq. (\ref{Regression}) alone
 (giving $r=0.83$, with a regression coefficient of $-0.000(2)$ for $Y_k(t_0)$).
Therefore, Eq. (\ref{Regression}) adequately captures output fluctuations that go beyond trivial sector size effects.
This confirms that the notion of economic susceptibility---the matrix $\rho^c_{ki}(t_0)$---coincides with (and is actually predictive of) the 
intuitive understanding that sectors with high susceptibility are those that are ``more easily moved'' by external events than low-susceptibility sectors.

We now show how the framework can be used to boost the quality of predictions of econometric timeseries models by extracting ``implied shocks'' from economic data.
Finally, we illustrate potential applications of our results by discussing estimates for economic impacts of recent tariffs imposed on US--EU trades in steel and aluminium. 

\subsubsection{Output predictions based on implied shocks}

The linear response formalism requires the specification of a demand shock in one or several sectors.
Such shocks, however, can rarely be observed directly in the data.
If a step demand shock occurs at the beginning of year $t$, the data from $t$ will not only contain the shock itself, 
but also of how the shock was ``digested'' by the economy during the year. 
As we have seen, recovery typically takes several years, see Fig. \ref{response}.
However, one can compute {\em implied shocks} from the data as follows (for clarity we omit the country index $c$ from now on).
Consider the ``truncated'' susceptibility matrix $\rho_{ik}(t,T)$, given by the area under the curve of the response function of $i$ to a shock in $k$, evaluated until $T$ years after the shock was applied, see Methods.
Assume that changes in output between year $t$ and $t+1$ are due to a step demand shock $\tilde X_i(t') = \theta(t'-t)\tilde X_i$, with $\theta$ the Heaviside step function, see Methods.
The size of this shock as implied by the output data from years $t$ and $t+1$ can be estimated by using Eq. (\ref{StepResponse}),
\begin{equation}
\tilde X_i = (\rho(t,T=1))_{ik}^{-1} \left( Y_k(t+1) -Y_k(t) \right)\quad.
\label{ImpliedShock}
\end{equation}
We refer to $\tilde X_i(t)$ as the {\em implied shock} at year $t$.
Positive (negative) output changes typically coincide with implied shocks that are of even larger (smaller) value, though some sectors defy these general trends, see SM Fig. S3. 

To test the validity of predictions of the linear response formalism, one can now take the implied shock from year $t$ and estimate the 
output in year $t+2$ using Eq. (\ref{StepResponse}). Note that, by construction, the output in year $t+1$ is identical in the model and the data.
This yields a LRT timeseries model for individual countries with a driven economy, 
\begin{equation}
\langle Y_k^{LRT}(t+2) \rangle_X =  Y_k(t) + \int_0^2 (\sigma^{-1})_{ij} \langle Y_k(t+\tau) Y_j(t) \rangle_0 \tilde X(t) d\tau \quad.
\label{LRTpure}
\end{equation} 
The predictions of the LRT timeseries model are compared with expectations from econometric timeseries forecasting methods, 
in particular to results from autoregressive integrated moving average (ARIMA) models \cite{Box70}; see SM Note S3 for a brief introduction.
The respective performance of the ARIMA and LRT model is evaluated by Pearson's correlation coefficient 
between the actual (empirically observed) and predicted output changes. 
For each year $t$ and country $c$, we compute the correlation coefficient $r^{LRT}(c,t)$, 
between the empirical output, $Y_k(t)$, and the predictions from the LRT model ($Y_k^{LRT}(t)$) in Eq. (\ref{LRTpure}).

Similarly, we compute the the correlation coefficients $r^{ARIMA}(c,t)$ for predictions from the ARIMA model 
(correlation of $Y_k^{ARIMA}(t)$ with $Y_k(t)$). 
Values for $r^{LRT}(c,t)$ and $r^{ARIMA}(c,t)$ for are shown in SM Figs. S4 and S5.

\begin{figure}[t]
\begin{center}
 \includegraphics[width = .5\textwidth, keepaspectratio = true]{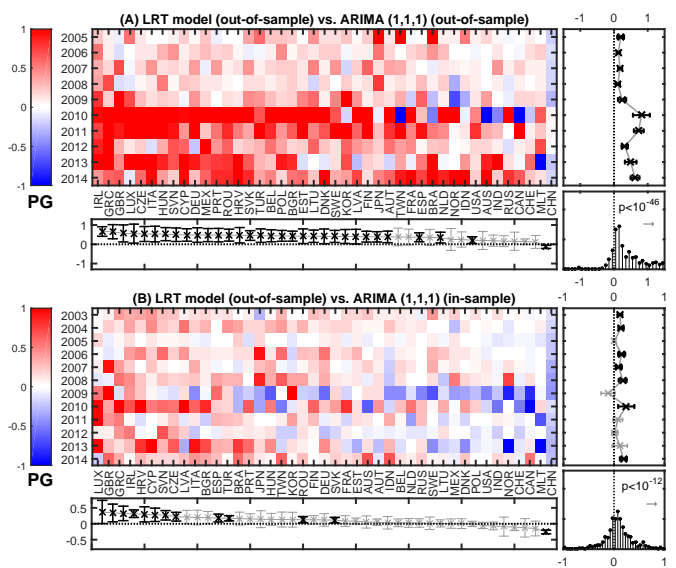}
\end{center}
 \caption{Comparison of the predictions of the linear response model with stochastic timeseries forecasting methods. 
 (A) Comparison of the LRT model for a shock between years $t$ and $t+1$ with an ARIMA(1,1,1) model that has been calibrated 
 using data up to year $t+1$. For every country and year, we show the  the  predictability gain, $PG$ of the LRT model over the ARIMA model.
In the panel to the right, $PG$ is averaged over all years, the bottom panel shows averages over countries. 
Averages that are significantly different from zero  are highlighted.  
The histogram of the $PG$ over all countries and years (bottom right) shows the corresponding distribution.
The LRT model drastically outperforms the ARIMA model, especially in the years that follow the crisis. 
The distribution of the predictability gains $PG$ over all countries and years is significantly skewed towards positive values ($p<10^{-46}$).
(B) As in (A), however, the ARIMA model is calibrated by using the complete information of the entire timeseries. 
Its predictions are still outperformed by the LRT model ($p<10^{-12}$).
This means that the out-of-sample predictions of the LRT model are superior to in-sample (!) predictions from standard econometric forecasting models.
Source data are provided as a Source Data file.
}
 \label{pg_pure}
\end{figure}

The differences between the correlation coefficients of two different models for the same country and year, is referred to as the ``predictability gain'',  $PG(c,t) = r^{LRT}(c,t)-r^{ARIMA}(c,t)$, see Fig. \ref{pg_pure}(A).
Red (blue) values indicate that for the given country and year the LRT model performs better (poorer) than the ARIMA model.
For every year, we perform a t-test to reject the null hypothesis that the true mean of $PG(c,t)$, taken over all countries, is zero ($p<0.05$).
The right panel in  Fig. \ref{pg_pure}(A) shows the $PG(c,t)$ averages over all countries taken at each year with a 95\% confidence interval (significant values are shown in black, non-significant in grey).
The bottom panels show the results for every country (significant vales are highlighted in black), and the histogram of $PG(c,t)$ taken over all years and countries.
The LRT model performs significantly better than the ARIMA model in almost each year and country.
We find predictability gains of up to 100\% and a $p$-value of $p<10^{-46}$ 
to reject the null hypothesis that the true mean of the distribution of $PG(c,t)$ is zero in this timespan.
Most intriguingly, for predictions from 2009 to 2010 (two years after the crisis occurred) the LRT model shows by far the largest predictability gains.
This result suggests that the LRT formalism works particularly well to describe the slow economic recovery during the Great Recession.

We design a further test, where it becomes harder for the LRT model to outcompete the ARIMA model, by comparing the 
out-of-sample predictions of the LRT model with the in-sample predictions of the ARIMA model. 
For this, we estimate the parameters of the sectoral ARIMA models over the entire timespan, from 2000 to 2014.
This should clearly stack the deck against the LRT model, as the ARIMA model is now calibrated using full timeseries information, 
in particular on the speed of economic recovery after the crisis.
Results are shown in Fig. \ref{pg_pure}(B). 
Overall, the LRT model again performs significantly better than the ARIMA model ($p<10^{-12}$).
The only exception is the prediction for 2009 where there is not clear which model is superior.
In this case the ARIMA model had the chance to ``learn'' the speed of the autoregression directly after the crisis.
In the following year, however, the LRT model shows the largest predictability gains, 
which again confirms that the LRT formalism is particularly useful to understand economic recovery.
Given that the ARIMA model has access to the full information of the timeseries, 
whereas the predictions of the LRT model are always taken entirely out-of-sample, 
these results once more confirm the superiority of the LRT formalism in describing the response of economies to large recessionary shocks.

Here we showed results for the ARIMA(1,1,1) model. However, qualitatively the same results 
are obtained (in many cases with even stronger relative performance of the LRT model) for other types of model. 
In particular, in SM we show results for the predictability gains $PG(c,t)$ for an ARIMA(1,1,0) model (differenced first order autoregressive model), an ARIMA(0,1,1) model (exponential smoothing), and an ARIMA(1,0,1) model (first order autoregressive moving average model) 
in Figs. S6, S7, and S8.
We also confirmed that the LRT model performs vastly superior to a sectoral VAR model, see SI Note S4 and Fig. S9.

\subsubsection{Indirect effects of the US--EU trade war}

\begin{figure}[t]
\begin{center}
 \includegraphics[width = .5\textwidth, keepaspectratio = true]{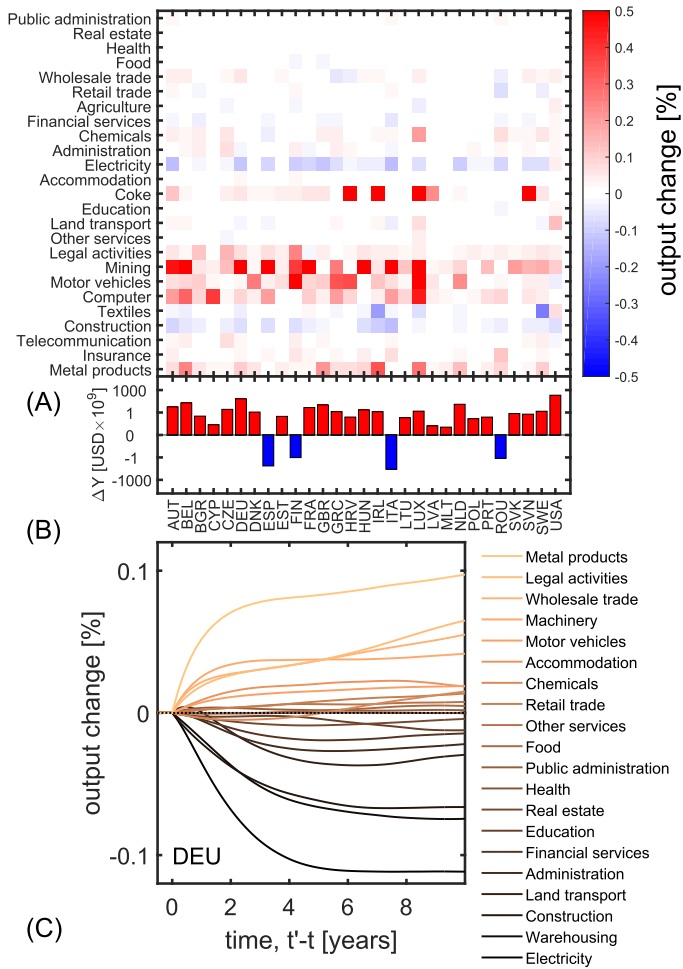}
\end{center}
 \caption{Estimation of indirect effects of the 2018 US steel and aluminum tariffs on EU countries. 
 (A) For all sectors and countries we estimate output-changes (in percent of 2014 outputs). Red (blue) colors indicate positive (negative) indirect effects.  Sectors that require basic metals as input (e.g. the manufacture of motor vehicles or fabricated metal products) tend to show positive indirect effects in Europe; negative in the US. On the other hand, sectors as electricity or wholesale trade show mostly negative impacts in the EU and positive ones in the US. 
 (B) For all countries we show the expected output change (in billion USD) due to indirect effects of the tariffs. Almost all countries experience positive indirect effects. Note that the $y$-axis scales logarithmically.
 (C) Response curves for Germany with a step demand shock in aluminum and steel.
 Source data are provided as a Source Data file.
}
 \label{tariffs}
\end{figure}

Finally we show how the LRT model can be used to estimate the economic impact of instances such as the currently escalating trade war 
between the EU and US \cite{Chu18}.
Starting from June 1, 2018, the US imposes a 25\% tariff on steel and a 10\% tariff on aluminum imports from member countries of the EU.
These tariffs are expected to lead to direct negative effects on EU steel and aluminum producers, 
which could be further amplified by other countries that redirect their exports from the US to the EU.
The indirect effects of these tariffs, however, are not so clear.
Increased supply of steel and aluminum in the EU might lead to a decrease in price with positive effects 
on industries that require those metals as inputs.
In the LRT model, the US tariffs impose a negative export demand shock on the manufacturing sector of basic metals (ISIC Code C24) on EU countries and a positive demand shock on the US.
We assume that US demand in this category will reduce by 100\% for European countries (and US domestic final consumption will increase accordingly) and estimate the resulting changes to the sectoral outputs using the linear relationship in Eq. (\ref{LinLaw}).
Note that the impacts of shocks with an arbitrary size of $x$\% of current export demand can simply be estimated by multiplying these results by $x/100$.
Results for $\langle \Delta Y_k \rangle_X$ obtained from Eq. (\ref{LinLaw}) using the most recent data available in WIOD ($t=2014$) are shown in Fig. (\ref{tariffs})(A) for the 25 largest sectors.
In general, output changes fall in the range between $\pm0.5$\%.
In European countries, positive effects are particularly strong in the manufacturing sectors (motor vehicles, computers, electronics, machinery, or electrical equipment), whereas there are consistently negative indirect effects for the energy sector.
These findings are consistent with an expectation of positive effects further down the supply chain of steel and aluminum (due to price decreases).
Decreases in the output of steel and aluminium production on the other hand coincide with a decrease in energy consumption.
It is also apparent that the indirect effects are distributed unequally across countries.
Manufacturing activities in Germany, Greece, or Ireland show consistently increased levels of output.
Indirect effects in the US often show opposite signs compared to the impact on European countries.
We find that negative indirect effects prevail for fabricated metal products and motor vehicles while the electricity sector, land transport, and wholesale trade experience positive effects. 
By summing the expected output-changes (in USD terms) over each sector in a country, we obtain the aggregated indirect effects; Fig. (\ref{tariffs})(B).
Overall, almost all countries experience positive indirect effects with output increases of up to several billion dollars; the exceptions being Spain, Finland, Italy, and Romania.
Our framework suggests that these countries might either (i) depend to a higher extent on sectors that provide input to the manufacture of basic metals (such as electricity), (ii) lack sectors that can profit from an increased supply of basic metals, or (iii) both of the former might be the case.  
Also, note that for European countries with positive aggregated indirect effects, these effects are typically outweighed by negative direct effects from the tariffs. 
Figure (\ref{tariffs})(C) shows the temporal impact (response curves) for Germany, for the a step demand shock for aluminum and steel.

\section*{Discussion}

We developed the theory of linear response for IO economies to quantify the resilience of national economies to production shocks.
We established an analytic link between stationary output fluctuations and their out-of-equilibrium behavior.
In particular, we derived the Green--Kubo relations for Leontief IO models, in full analogy to a wide range of physical phenomena, 
ranging from electrical and magnetic susceptibilities to shear viscosity and electrical resistance.
Our framework can be applied to other types of IO model, as long as they are (i) linear and (ii) permit a stationary solution.
This includes IO models that use a higher geographic resolution (i.e. regional IO models), 
but also several of their generalizations, such as environmentally extended IO models \cite{Wiedmann13}, 
or commodity-by-industry IO models \cite{Miller09}.

The central result of our work is a linear ``phenomenological law'' between demand shock, $X_i$, and the induced output change $\Delta Y_k$, namely that 
$\langle \Delta Y_k \rangle_X = \rho_{ki} X_i$, with $\rho$ being a sector-by-sector matrix of economic susceptibilities.
The output change $\langle \Delta Y_k \rangle_X$  characterizes a driven economy in a nonequilibrium stationary state.
The original equilibrium state, $(\mathbb I -\mathbf{A})^{-1}\mathbf{D}$, is recovered for $X_i(t)=0$ for all $i$ and $t$.
The LRT solution $\langle \Delta Y_k \rangle_X$ is also fundamentally different from the ``perturbed equilibrium state'' implied by a step demand shock of the form $\mathbf{D_P} = \mathbf{D}+\mathbf{X}$ with $\mathbf{X}(t) =\mathbf{X} \theta(t)$, 
namely the perturbed equilibrium state  $(\mathbb I -\mathbf{A})^{-1}\mathbf{D_P}$.
To see the difference, note that in LRT the expectation values are taken over the probability density function of the stationary solution of $\mathbf{\dot Y} = (\mathbf{A}-\mathbb I)\mathbf{Y} + \mathbf{D} + \mathbf{F}(t)$ (from which the non-equilibrium expectation value $\langle \Delta Y_k \rangle_X$ is estimated), 
whereas the perturbed equilibrium state would be given by expectation values using the stationary solution of $\mathbf{\dot Y} = (\mathbf{A}-\mathbb I)\mathbf{Y} + \mathbf{D_P} + \mathbf{F}(t)$ as probability measure; see also SM Note S6 and Fig. S10.

We demonstrated that our measures for economic susceptibility that can be derived from data, 
are indeed predictive of future output fluctuations, even when no knowledge of future shocks is available.
This finding corroborates that sectors with high susceptibility are indeed those that tend to be 
``more easily movable'' by external shocks than low-susceptibility sectors.
We showed that out-of-sample predictions from the LRT model consistently outperform standard econometric forecasting methods, 
such as different types of ARIMA model.
Predictions of the LRT model work particularly well in the years that followed the recent financial crisis.
This suggests that the LRT formalism allows us to get an analytic and quantitative understanding 
of the slow economic recovery of certain countries in the wake of the Great Recession.
Because of the versatility and conceptual simplicity of input--output models, our framework can lead to more accurate quantitative estimates for the impact of disruptive events in various applications and scales, ranging from global recessions to regional, critical infrastructure systems.
We illustrate the practical usefulness of our approach in providing concrete estimates 
for the indirect effects of the currently escalating US--EU trade war.
In particular, we considered a negative export demand shock on the manufacturing sector of basic metals on EU countries and a corresponding positive demand shock on the US.
We find that in European countries there is a trend toward positive indirect effects for manufacturing sectors further down the supply chain from basic metals, whereas electricity outputs show negative indirect effects.
In the US we find similar results with reversed signs; positive (negative) effects moving further up (down) along the supply chain.

A limitation of the Leontief IO model that extends to our work is that prices play no role in the model.
Firms in real economies can respond to shocks by adjusting produced quantities as well as prices.
It therefore remains to be seen how prices can be incorporated in the LRT framework, i.e., within a linear time-invariant formulation of the underlying microscopic dynamics.
Besides linearity and time-invariance, our approach also assumes an external shock that may depend only on time and for which we only consider first order correction with respect to the unperturbed state.

In summary, in this work we extended current mainstream economic theories to out-of-equilibrium situations 
in a way that is analytically rigorous, empirically testable, and flexible enough to immediately address a wide range of scenarios with a direct political relevance, such as identifying those parts of a country's economy that are particularly vulnerable in a trade war.

\section*{Methods}

\subsection{Linear response theory of input-output economics}  

Consider an economy with $N$ sectors, each sector producing $Y_i$ units of a single homogeneous good.
Assume that sector $j$ requires $A_{ij}$ units from sector $i$ as input to produce one unit itself, giving the so-called technical coefficients $A_{ij}$.
Each sector sells some of its output to consumers, the demand $D_i$.
The open Leontief IO model, {\em the} standard model in economics to depict and analyze inter-sectoral relationships, 
assumes linear production functions given by $\mathbf{Y} = \mathbf{AY}+\mathbf{D}$ (matrix notation).  
The stationary (equilibrium) state of this economy is given by $\mathbf{Y^0}=(\mathbb I -\mathbf{A})^{-1}\mathbf{D}$ ($\mathbb I$ being the $N$-dimensional identity matrix).
For the time evolution of an economy in its stationary state, 
assuming that differences in dynamic demand $\mathbf{AY}+\mathbf{D}$ and dynamic production $\mathbf{Y}$ are compensated by 
production changes, this model gives the differential equation \cite{Leontief86}
\begin{equation}
\mathbf{\dot Y} = (\mathbf{A}-\mathbb I)\mathbf{Y} + \mathbf{D} \quad.
\label{LIOM}
\end{equation}
We assume (i) that each sector $i$ experiences a time-dependent demand shock, $X_i(t)$, and (ii) the presence of multivariate white noise, i.e. a stochastic force, $F_i(t)$.
In the picture of the example of the house of cards given above, the noise $F_i(t)$ represents the ``tiny nudge'' that we apply to the serving tray to understand if the house would survive a much larger shock, $X_i(t)$.
More formally, the nudge consists of noise with mean value $\langle F_i(t) \rangle_0=0$, 
and covariance $\langle F_i(t) F_j(s) \rangle_0=\nu_{ij} \delta(t-s)$.
Here, $\delta(x)$ denotes the Dirac-delta function, $\nu$ is a matrix of constants, and $\langle \mathbf{x}(t) \rangle_0 = \int d^N\mathbf{Y} \mathbf{x}(\mathbf{Y}) f_0(\mathbf{Y})$
 is the {\em equilibrium expectation value} of the function $\mathbf{x}(t)$, evaluated in the absence of an external force ($\mathbf{X}(t)=0$), with $f_0(\mathbf{Y})$ being the probability distribution to find a given value of $\mathbf{Y}$ under noise $F_i(t)$.
This leads to the stochastic differential equation, 
\begin{equation}
\mathbf{\dot Y} = (\mathbf{A}-\mathbb I)\mathbf{Y} + \mathbf{D} + \mathbf{X}(t) + \mathbf{F}(t) \quad.
\label{SLIOM}
\end{equation}
From the central limit theorem it follows immediately that the stationary or equilibrium solution $f_0(\mathbf{Y})$ in the absence of external shocks ($\mathbf{X}(t)=0$) of Eq. (\ref{SLIOM}) is given by a multivariate normal distribution with covariance $\sigma_{ij} = \lim_{t\to\infty} \langle Y_i(t) Y_j(t) \rangle_0$.
In the presence of external shocks, i.e., for $\mathbf{X}(t) \neq 0$, a solution of Eq. (\ref{SLIOM}) with first order corrections from the shock can be obtained using LRT, see SM Note S5.
We denote the expectation value for the output change of sector $k$ with nonzero shock $\mathbf{X}(t)$ by $\langle \Delta Y_k(t) \rangle_X \equiv \langle Y_k(t) \rangle_X-Y_k^0$.
That is, averages with a subscript $0$ refer to values taken {\em at equilibrium}, 
whereas averages with a subscript $X$ refer to {\em out-of-equilibrium} properties.
Following LRT \cite{Evans90}, we get the general solution for the time evolution of the output changes, $\langle \Delta Y_k(t) \rangle_X$,
\begin{equation}
\langle \Delta Y_k(t) \rangle_X = \int_{-\infty}^t (\sigma^{-1})_{ij} \langle Y_k(\tau) Y_j(0) \rangle_0 X_i(\tau) d\tau \quad. 
\label{GenResponse}
\end{equation}
Remarkably, we have related the out-of-equilibrium response of the sectoral outputs, 
$\langle \Delta Y_k(t) \rangle_X$, to their correlation functions taken {\em at} equilibrium.
Equation \ref{GenResponse} characterizes the state of a {\it driven} economy.

For certain types of demand shock, the resulting output change takes a particularly simple form.
For an impulse demand shock, $X_i(t) = \delta(t) X_i$, we get 
\begin{equation}
\langle \Delta Y_k^{pulse}(t) \rangle_X = (\sigma^{-1})_{ij} \langle Y_k(t) Y_j(0) \rangle_0 X_i \quad. 
\label{PulseResponse}
\end{equation}
For a step demand shock, $X_i(t) = \theta(t) X_i$ with the Heaviside step function $\theta(t \geq 0)=1$ and $\theta(t<0)=0$, we get 
\begin{equation}
\langle \Delta Y_k^{step}(t) \rangle_X = \int_0^t (\sigma^{-1})_{ij} \langle Y_k(\tau) Y_j(0) \rangle_0 X_i d\tau \quad. 
\label{StepResponse}
\end{equation}
For $t\gg0$ we obtain the linear relation
\begin{equation}
\langle \Delta Y_k  \rangle_X = \rho_{ki} X_i, \ \mathrm{with} \ \rho_{ki} = \int_0^{\infty} (\sigma^{-1})_{ij} \langle Y_k(\tau) Y_j(0) \rangle_0 d\tau \quad, 
\label{LinLaw}
\end{equation}
where we introduced the economic susceptibility $\rho$,  
in full analogy to the derivation of electric or magnetic susceptibilities in statistical mechanics, see SM Table 1.
The economic susceptibility $\rho_{ki}$ has the precise meaning of output-change in sector $k$, given that a step demand shock of unit size occurs in sector $i$.  

In this paper we encounter different types of susceptibility, depending on how averages are taken. 
In particular we will use the following definitions: 
The $N \times N$ susceptibility matrix of a country $c$ at year $t$ is defined by 
\begin{equation}
	\rho_{ij}^c(t)=  \int_0^{\infty} (\sigma^{-1})_{ij} \langle Y_k^c(t+\tau) Y_j^c(t) \rangle_0 d\tau \quad,
	\label{RhoBasic}
\end{equation} 
where the output $Y_k^c(t)$, technical coefficients $A_{ij}^c(t)$, and the demand $D_i^c(t)$ of a particular country $c$, 
are read off the WIOD \cite{Timmer15}.
A truncated version $\rho_{ij}^c(t,T)$ of this susceptibility matrix is obtained by taking $t+T$, $T>0$, as the upper boundary of the integration range in Eq. (\ref{RhoBasic}).
The susceptibility of a sector $i$ in country $c$ at year $t$, $\rho_i^c(t)$, is defined as the corresponding column sum of the susceptibility matrix, $\rho_i^c(t) = \sum_j \rho_{ij}^c(t)$.
We define the {\em averaged country susceptibility} as the average of the sector susceptibility taken over all $N$ sectors and $N_t=15$ years, 
$\rho^c = (N_tN)^{-1} \sum_{i,t} \rho_i^c(t)$. 
The {\em output-weighted average sector susceptibility}, $\rho_i$ is defined as, 
$\rho_i = (N_t N_c \sum_{t,c} Y^c_i(t))^{-1} \sum_{t,c} Y^c_i(t) \rho_i^c(t)$, where $N_c=43$ is the number of countries in the data.

\subsection{Acknowledgments}  
We thank M. Miess and A. Pichler for helpful discussions, Johannes Sorger for help with the visualizations, and acknowledge support from the European Commission, H2020 SmartResilience No. 700621, FFG Project 857136, and OeNB  Jubil\"aumsfond project 17795.

\subsection{Author contributions}
PK and ST designed research, PK performed research and analyzed data, SP contributed timeseries models, PK and ST wrote the paper.

\subsection{Conflict of interest}
The authors declare no conflict of interest.

\subsection{Data availability}
The study is based on the 2016 release of the World Input--Output Tables \cite{Timmer15} (see also \href{http://www.wiod.org}{http://www.wiod.org}, accessed 15 January 2019) .

\subsection{Code availability}
Code is available upon request directly from the authors.


\begin{thebibliography}{99}

\bibitem{IMF09} International Monetary Fund, World Economic Outlook 2009, Crisis and Recovery (International Monetary Fund, Washington, DC, USA, 2009).

\bibitem{Barnett14} Barnett A, Batten S, Chiu A, Franklin J, Sebasti\'a-Barriel, The UK productivity puzzle. {\it Bank of England Quarterly Bulletin} {\bf 54}(2), 114--28, 2014.

\bibitem{Bryson14} Bryson A, Forth J,  Askenazy P, Productivity Puzzles in Europe: a Comparison of the UK, France, Germany and Spain. NIESR Discussion Paper {\bf 448}, 2014.

\bibitem{Schweitzer09} Schweitzer F, Fagiolo G, Sornette D, Vega-Redondo F, Vespignani A, White D, Economic networks: the new challenges. {\it Science} {\bf 325}, 422--25, 2009.

\bibitem{Garas10} Garas A, Argyrakis P, Rozenblat C, Tomassini M, Havlin S, Worldwide spreading of economic crisis. {\it New Journal of Physics} {\bf 12}, 113043, 2010.

\bibitem{Haldane11} Haldane A, May R, Systemic risk in banking ecosystems. {\it Nature} {\bf 469}, 351--5, 2011.

\bibitem{Farmer12} Farmer JD, Gallegati M, Hommes C, Kirman A, Ormerod P, Cincotti S, Sanchez A, Helbing D, A complex systems approach to constructing better models for managing financial markets and the economy. {\it Eur. Phys. J. Special Topics} {\bf 214}, 295--324 (2012).

\bibitem{Martin11} Martin R, Regional economic resilience, hysteresis and recessionary shocks. {\it Journal of economic geography} {\bf 12}(1), 1--32, 2011.

\bibitem{Kirman10} Kirman A, The economic crisis is a crisis for economic theory. {\it CESifo Economic Studies} {\bf 56}(4), 498--35, 2010.

\bibitem{Arrow54} Arrow KJ, Debreu G, Existence of an equilibrium for a competitive economy. {\it Econometrica} {\bf 22}(3), 265--90, 1954.

\bibitem{Smets07} Smets F, Wouters R, Shocks and frictions in US business cycles: A Bayesian DSGE approach. {\it The American Economic Review} {\bf 97}(3), 586--606, 2007.

\bibitem{Mitra08} Mitra-Kahn BH, Debunking the myths of computable general equilibrium models. SCEPA Working Paper 2008-1, 2008.

\bibitem{Farmer05} Farmer JD, Smith DE, Shubik M, Is economics the next physical science? {\it Physics Today} {\bf 58}(9), 37--42, 2005.

\bibitem{Onsager31} Onsager L, Reciprocal relations in irreversible processes. I. {\it Physical Review} {\bf 37}(4), 405, 1931.

\bibitem{Kubo66} Kubo R, The fluctuation-dissipation theorem. {\it Reports on progress in physics} {\bf 29}(1), 255, 1966.

\bibitem{Nicolis77} Nicolis G, Prigogine I, Self-organization in nonequilibrium systems (Wiley, New York, USA, 1977).

\bibitem{Hanel14} Hanel R, Thurner S, Gell-Mann M, How multiplicity determines entropy and the derivation of the maximum entropy principle for complex systems. {\it Proceedings of the National Academy of Sciences USA} {\bf 111}(19), 6905--10, 2014.

\bibitem{Jarzynski15} Jarzynski C, Diverse phenomena, common themes. {\it Nature Physics} {\bf 11}(2), 105--7, 2015.

\bibitem{Gallegatti06} Gallegatti M, Keen S, Lux T, Ormerod P, Worrying trends in econophysics. {\it Physica A} {\bf 370}(1), 1--6, 2006.

\bibitem{Farmer09} Farmer JD, Foley D, The economy needs agent-based modelling. {\it Nature} {\bf 460}, 685--6, 2009.

\bibitem{Klimek14} Klimek P, Poledna S, Farmer JD, Thurner S, To bail-out or to bail-in? Answers from an agent-based model. {\it Journal of Economic Dynamics and Control} {\bf 50}, 144--54, 2014.

\bibitem{Evans90} Evans DJ, Morriss GP, Statistical mechanics of nonequilibrium liquids (Theoretical Chemistry Monograph Series, Academic Press, London, UK, 1990).

\bibitem{Leontief86} Leontief W, Input-output economic (Oxford University Press, New York, USA, 1986).

\bibitem{Green54} Green MS, Markoff random processes and the statistical mechanics of time-dependent phenomena. II. Irreversible processes in fluids.  {\it J. Chem. Phys} {\bf 22}, 398--413, 1954.

\bibitem{Kubo57} Kubo R, Statistical-mechanical theory of irreversible processes. I.
General theory and simple applications to magnetic and conduction problems. {\it J. Phys. Soc. Jpn.} {\bf 12}, 570--586, 1957. 

\bibitem{Timmer15} Timmer MP, Dietzenbacher E, Los B, Stehrer R, de Vries GJ, An illustrated user guide to the world input-output database: the case of global automotive production. {\it Review of International Economics} {\bf 23}, 575--605, 2015.

\bibitem{Long83} Long JB, Plosser CI, Real Business Cycles. {\it Journal of Political Economy} {\bf 91}(1), 39--69, 1983.

\bibitem{Haimes01} Haimes YY, Jiang P, Leontief-based model of risk in complex interconnected infrastructures. {\it Journal of Infrastructure Systems} {\bf 7}(1), 1--12, 2001.

\bibitem{Contreras14} Contreras MGA, Fagiolo G, Propagation of economic shocks in input--output networks: A cross-country analysis. {\it Physical Review E} {\bf 90}(6), 062812, 2014.

\bibitem{Acemoglu16} Acemoglue D, Akcigit U, Kerr W, Networks and the macroeconomy: An empirical exploration. {\it NBER Macroeconomics Annual} {\bf 30}(1), 273--335, 2016.

\bibitem{He17} He P, Ng TS, Su B, Energy-economic recovery resilience with input--output linear programming models. {\it Energy Economics} {\bf 68}, 177--191, 2017.

\bibitem{Phillips13} Phillips CL, Parr JM, Riskin EA, Signals, systems, and transforms (Prentice Hall, 2013). 

\bibitem{Pesaran98} Pesaran HH, Shin Y, Generalized impulse response analysis in linear multivariate models. {\it Economic letters} {\bf 58}(1), 17--29, 1998.

\bibitem{Lutkepohl08} L\"utkepohl M, Impulse response function. In: Palgrave Macmillan (eds) The New Palgrave Dictionary of Economics (Palgrave Macmillan, London, 2008).

\bibitem{Serrano09} Angeles Serrano M, Boguna M, Vespignani A, Extracting the multiscale backbone of complex weighted networks. {\it Proceedings of the National Academy of Sciences USA} {\bf 106}(16), 6483-8, 2009.

\bibitem{Timmer16} Timmer M, Los B, Stehrer R, de Vries G, An anatomy of the global trade slowdown based on the WIOD 2016 release (No. GD-162) (Groningen Growth and Development Centre, University of Groningen, 2016). 

\bibitem{Box70} Box GE, Jenkins GM, Reinsel GC, Ljung GM, Time series analysis: forecasting and control (John Wiley \& Sons, New Jersey, USA, 1970).

\bibitem{Chu18} Chu B, How economically damaging will Trump's steel tariffs be? The Independent, 31 May 2018. (\href{https://www.independent.co.uk/news/business/analysis-and-features/trump-us-steel-tariff-eu-damaging-explained-trade-war-uk-jobs-a8377871.html}{https://www.independent.co.uk}, accessed 18 July 2018).

\bibitem{Wiedmann13} Wiedmann TO, Schandl H, Lenzen M, Moran D, Suh S, West J, Kanemoto K, The material footprint of nations. {\it Proceedings of the National Academy of Sciences USA} {\bf 112}(20), 6271-6, 2013.

\bibitem{Miller09} Miller RE, Blair PD, Input-output analysis: foundations and extensions (Cambridge University Press, Cambridge, UK, 2009).



\end{thebibliography}

\newpage

\section*{Supplementary Material}

\subsection{Note S1:  General information on linear response theory}

The basic idea of LRT is to compute the explicit time evolution of an external shock to a system in equilibrium. 
Originally, LRT has been formulated with thermal and mechanical transport processes in mind.
The idea is that a thermodynamic system (e.g. a canister of gas, a block of metal, a piece of lumber, ...) can be prevented from relaxing to an equilibrium configuration by an external field.
This field acts either directly on the system's microscopic constituents (think of an electric or magnetic field influencing the electrons in a block of metal) or that acts on the system boundaries (e.g. by generating a temperature gradient or by deforming the container of a system).
In both situations the external field drives the system from equilibrium by means of induced transport processes.
An electric field induces a current (e.g. transport of electrons) while a temperature gradient induces a heat flow (transport of particles with high kinetic energy).
The central result of LRT is that a specific current (or flux) can be associated with such an external field.
Furthermore, the magnitude of this flux is proportional to the magnitude of the external field.

Melville S. Green \cite{Green54} and Ryogbo Kubo \cite{Kubo57} independently found the exact mathematical expression that relates the field to its induced flux, the so-called transport coefficients $\rho$.
These expressions are now called Green--Kubo relations. 
Note that $\rho$ can be a scalar, a matrix, or a higher dimensional tensor, see Table S1.
Their point of departure was that thermodynamic systems undergo a certain amount of fluctuations even at equilibrium.
Given a perturbation in one of the microscopic degrees of freedom of the system (e.g. a fluctuation in a particle's velocity along a certain direction), the microscopic equations of motion allow to compute how strong and at which time other degrees of freedom (other particles) will ``feel'' this fluctuation.
LRT assumes that other, larger perturbations will be felt in a similar way, i.e. with a similar delay and attenuation.
Note that in many cases it is not even remotely possible to get a firm analytical understanding of the thermodynamics of a system once it is removed from equilibrium, given that it makes sense to talk about thermodynamics in such regimes at all.
LRT provides one of the few analytic links that allows us to extrapolate out-of-equilibrium properties of a system from its behavior in equilibrium and therefore one of the paramount achievements of 20th century statistical mechanics.

\subsection{Note S2: Description of the World Input--Output Database}

In this work we use the World Input--Output Database (WIOD) release from November 2016 (http://www.wiod.org/release16).
This release consists of several data tables that cover 28 EU countries and 15 other major countries (Australia, Brazil, Canada, China, India, Indonesia, Japan, Mexico, Norway, Russia, South Korea, Switzerland, Taiwan, Turkey, and the United States) from 2000 to 2014.
The 15 non-European countries were chosen in order to cover all major parts of the world economy while ensuring that the data is available at a sufficient quality \cite{Timmer16}.
The resulting 43 countries account for more than 85\% of world GDP \cite{Timmer16}.
For each year and country, WIOD contains data on 56 sectors according to the 2-digit ISIC revision 4 level.

The basic construction steps of the WIOD are as follows \cite{Timmer16}.
First, timeseries of national supply and use tables are constructed for each country based solely on national account statistics.
With the use of international trade databases, these timeseries are disaggregated into imports by country of origin and use category.
Finally, the resulting data items are integrated into a single input--output table.
The construction of WIOD uses solely national account statistics as input for the construction of national supply and use tables.
This ensures that (i) WIOD is fully consistent with national accounts and (ii) the use of a consistent methodology in categorizing products and services across time.
This comes at the cost of a smaller coverage of countries and years with respect to other comparable databases that do not ensure such consistencies, see also \cite{Timmer15} and \cite{Timmer16}.

\subsection{Note S3: A brief introduction to ARIMA models}

ARIMA models are a quite general class of models for stationary timeseries, that is for timeseries that are characterized by statistical properties that do not change over time \cite{Box70} (we only consider nonseasonable models here).
Stationary timeseries show variations around their means that are of constant amplitude and that look similar over time.
That is, given a snippet from a timeseries it is impossible to deduce when exactly the snippet was observed, e.g. after a couple of time steps or after observing the process for a million years.
ARIMA models serve the purpose of forecasting timeseries based on observations of previous values that were collected within a specific time interval.
It turns out that stationary timeseries have a quite limited set of properties that fully characterize them.  
These properties are encoded in the acronym ARIMA, which stands for auto-regressive integrated moving average.
Auto-regression (AR) means that the next value of a timeseries can be predicted as a multiple of its prior values (maybe plus a constant term).
Integration (I) means the timeseries should be forecasted by considering (also) {\em differences} between prior values, rather than (only) the values themselves.
Moving average (MA) finally indicates that the timeseries is best predicted by considering not the last observed value (as in random walks without memory), but rather the average over several previously observed values.
Each of these three properties, AR, I, and MA, can be present up to a specific order.
For AR this order is the number $p$ of auto-regressive terms, for I the number $d$ of differences that are relevant, and for MA it is the number $q$ of past observations that are included in the average.
An ARIMA model is specified by the choice of these three numbers as an ARIMA($p$,$d$,$q$) model. 
If the number of observations available to calibrate the ARIMA model is low (as it is the case for our work), it is often not meaningful to consider orders of parameters higher than one or two, as the corresponding coefficients for the higher order correction terms can not reliably be estimated.

\subsection{Note S4: Comparison of LRT and VAR models}
Vector autoregression (VAR) models can be used to describe stochastic processes driven by linear dependencies on multiple other variables.
In our case we can consider the sectoral outputs $Y_k^c(t)$ of country $c$ as a stochastic process that may depend on the output of all other $N$ sectors in the same country at the previous timestep, the outputs $Y_j^c(t-1)$ with $j\in\{1,\dots,N\}$.
Note that ARIMA models assume that output changes in each sector can be predicted based on past values of the same sector only.
By benchmarking the LRT model against a VAR model we can go beyond this limitation and test the LRT framework against a regression model that captures the structure of inter-industry dependencies in a more comprehensively way.
The drawbacks of VAR models are that they do not scale well with system size.
For a first order VAR model with $N=56$ sectors one needs to estimate no less than $N^2=3,136$ different parameters.
It is therefore completely hopeless to specify a sectoral VAR model from data given that we have only 15 observations for each sector.
However, we can use a strategy to calibrate a sectoral VAR model similar to how we measured the response functions in the LRT framework.
By assuming that the economy can be represented by the stochastic differential equation,
\begin{equation}
\dot Y = (A-\mathbb I)Y + D  + F(t)\quad,
\label{SLIOM2}
\end{equation}
we generated 10,000 synthetic observations for the output of each sector.
Using supercomputing resources (the Vienna Scientific Cluster 3, one of the hundred fastest computers worldwide (http://vsc.ac.at/systems/vsc-3/, accessed Sep 14, 2018.) we were then able to estimate the parameters of a first order VAR model using data from one year ($t=2000$) for each country.
By construction, the model has no linear trend.
It remains to estimate the entries of the autoregression matrix $AR^c$ for each country $c$, given by $Y_k^c(t+1) = AR^c Y_k^c(t) + e_k^c$ ($e_k^c$ being the intercepts).
We then compare the forecasts of this VAR models with predictions of the LRT model, similar to the comparison with ARIMA models.
The results are shown in Fig. S7.
For almost all years and countries, the LRT model performs vastly superior than the VAR model ($p<10^{-117}$).
The advantage of the LRT model is least pronounced in the years 2009 and 2010 (where LRT still performs significantly better), whereas other years show predictability gains in the range between one and two.
Note that these results are based on a VAR model calibrated in 2000 only due to the costly requirement for supercomputing resources.
For a single country (Germany), we evaluated the VAR model also for several years afterwards and in particular for the crisis year 2008.
Still, we found qualitatively the same result as shown in Fig. S7 with the majority of predictability gains lying in the range between one and two.

\subsection{Note S5: Derivation of Green--Kubo relations for input--output economics} \label{S2}

Let $y \equiv \{ y_i(t) \}, i=1,\dots,N$ be the stochastic variables describing the time evolution of a dynamical system.
In the following we assume the system to be linear and time invariant.
Linearity means that the time evolution of the system is governed by a linear operator.
Time invariance means that the response of the system does not depend on the time at which we apply the shock.
We describe the state of the system by its probability density function, $f(y,t)$.
The time evolution of the system is described by a linear time operator, $L_0(y)$, for instance a Fokker-Planck operator.
In general, $L_0(y)$ can be any linear operator with a stationary solution for $f(y,t)$ for the LRT framework to be applicable.
Further, the system is perturbed by a time-depending external field $X(t)$, i.e. we allow the external field to vary arbitrarily over time but not as a function of the stochastic variables $y$..
Under perturbation, the time evolution of $f(y,t)$ is then described by a perturbed operator $\Lambda_{X}(y,t)$, as
\begin{eqnarray}
\Lambda_{X}(y,t) & = & L_0(y) + L_X (y,t) \quad, \\
L_X(y,t) & = & L_X(y) X(t) \quad,
\label{XOperator}
\end{eqnarray}
such that 
\begin{equation}
\frac{\partial f}{\partial t} = L_{0}(y,t) f(y,t) \quad.
\label{FPE}
\end{equation}
Let us denote the stationary solution for $L_{0}$ by $f_0(y)$, given by $L_0(y) f_0(y) = 0$.

\subsubsection{Response Function}

Consider a small perturbation of the stationary solution of the form $f(y,t) = f_0(y) + \Delta f(y,t)$.
For the time evolution we have then
\begin{eqnarray}
\dot{f}(y,t) & = & \Delta \dot{f}(y,t) = \Lambda_{X}(y,t) f(y,t) \\
  & = & \left( L_0(y) +L_X(y,t) \right) \left( f_0(y) + \Delta f(y,t) \right)  \\
  & = & L_0(y) \Delta f(y,t) + L_X(y,t) f_0(y) \quad. \label{approxim}
\end{eqnarray}
The above equation can be solved using well-known properties of the Laplace transformation, given by $\tilde f (s) = \int_0^t dt \mathrm{e}^{-st}f(t)$.
For the l.h.s. of the equation  $\dot{f}(y,t) = L_0(y) \Delta f(y,t) + L_X(y,t) f_0(y)$ we get,
\begin{equation}
\dot{f}(y,t) = \Delta \dot{f}(y,t) = s \Delta \tilde f(s) - \Delta f(0) \quad,
\end{equation}
where we can use that $\Delta f(0) =0$.
The r.h.s. simply transforms to $L_0(y) \Delta f(y,t) + L_X(y,t) f_0(y) = L_0(y) \Delta \tilde f(s) + f_0(y) \tilde L_X$.
We obtain, 
\begin{equation}
 \Delta \tilde f (s) = \frac{\tilde{L_X} f_0(y)}{s-L_0(y)} \quad.
 \label{LaplTr}
 \end{equation}
 Now we apply an inverse Laplace transformation to the l.h.s. and r.h.s. of Eq. \ref{LaplTr}.
 Therefore, we make use of the Laplace transform properties that (i) for a product of functions we have $\tilde f(s) \tilde g(s) =\int d\tau f(\tau) g(t-\tau)$ and (ii) that the exponential function, $\mathrm e^{at}$ transforms to $1/(s-a)$.
 Using these properties, Eq. \ref{LaplTr} becomes
\begin{equation}
 \Delta f(y,t) = \int_{-\infty}^t d\tau \ \mathrm e^{L_0(y) \cdot (t-\tau)} L_X(\tau) f_0(y) \quad.
 \label{solu}
\end{equation}

We are now interested in the expectation value of any dynamical variable $B(t)$ under the considered perturbations.
Formally, we have
\begin{eqnarray}
\langle B(t) \rangle & = & \int d^N y B(y) f(y,t) \\
 & = & \langle B \rangle_0 + \int d^N y B(y) \Delta f(y,t) \quad.
\end{eqnarray}
By plugging in the solution for $\Delta f(y,t)$ obtained in Eq. \ref{solu} we get
\begin{equation}
\langle B(t) \rangle =  \langle B \rangle_0 + \int _{-\infty}^t R_{B,X} (t-\tau) X(\tau) d \tau \quad,
\label{ExpVal}
\end{equation}
where $R_{B,X}(t)$ is the {\it response function}
\begin{eqnarray}
R_{B,X}(t) & = &  \int d^N y \ B(y) \mathrm e^{L_0(y) t} L_X(y) f_0(y)  \   t\geq 0, \nonumber \\
R_{B,X}(t) & = & 0 \quad \text{for } t< 0.
\label{RespFunc}
\end{eqnarray}
$R_{B,X}(t)$ describes the response of the variable or observable $B(t)$ to the external force $X(t)$.

\subsubsection{Correlation Functions}

The response function can be expressed in terms of equilibrium correlation functions $c_{A,B}(\tau)$ between an observable $A(y(t))$ and $B(y(t))$ with lag $\tau$.
First, observe that
\begin{eqnarray}
c_{A,B}(\tau) & = & \langle A(y(\tau)) B(y(0)) \rangle \\
 & = & \int d^Ny \ d^N y_0 \ A(y) B(y_0) P(y,\tau; q_0,0) \quad, \nonumber 
\end{eqnarray}
where $P(y,\tau; y_0,0)$ is the joint probability that the system is in configuration $y_0$ at time $t=0$ and in configuration $y$ at $t$.
This can be written as 
\begin{equation}
P(y,\tau; y_0,0) = P(y,\tau| y_0,0) P(y_0,0) \quad.
\end{equation}
Here, $P(y,\tau| y_0,0)$ is the conditional probability that the system evolved from state $y_0$ at $t=0$ to state $y$ at $t$.
Formally, this conditional probability is equal to the propagator $\mathrm e^{L_0(y)\cdot \tau} \delta(y-y_0)$.
Note that $P(y_0,0)$ is the stationary solution $f_0(y_0)$.
For the equilibrium correlation function this means that
\begin{equation}
 c_{A,B}(\tau) = \int d^Ny \ A(y) \mathrm e^{L_0(y)\cdot \tau} B(y) f_0(y) \quad.
\end{equation}
Define the function $A(y)$ as $A(y) \equiv f_0^{-1}(y) L_X(y) f_0(y)$ and introduce the {\it generalized potential} $\phi(y)$ through the relation
\begin{equation}
f_0(y) \equiv \mathcal N \mathrm e^{- \phi(y)} \quad,
\end{equation}
with $\mathcal N$ as normalization constant.
With these definitions we re-write the correlation function as
\begin{equation}
c_{B,A}(\tau) =  \int d^N y \ B(y) \mathrm e^{L_0(y) t} L_X(y) f_0(y) \quad,
\end{equation}
and obtain the result that
\begin{equation}
c_{B,A}(\tau) =  R_{B,X} (\tau) \quad.
\end{equation}

\subsubsection{Application to stochastic processes}

Consider a stochastic dynamical system with an external force $X_i(t)$ acting on component $i$ and the stochastic force $F_{R,i}(t)$ with the properties 
$\langle F_{R,i}(t) \rangle = 0$ and $\langle F_{R,i}(t)\cdot F_{R,j}(t') \rangle = \epsilon_{ij} \delta(t-t')$.
In particular, we consider linear systems of the type,
\begin{equation}
\dot y_i + \sum_{j=1}^N \gamma_{ij}y_j = F_{R,i}(t)+X_i(t) \quad.
\end{equation}
Clearly, the stochastic IO dynamics that we consider in the main text is of the above type.
To this end, note that one can always use a variable transformation of the form $Y \to y = Y - (\mathbb I - A)^{-1}D$ to rewrite the IO model into a homogeneous differential equation.
From now on, we will use the summation convention.
The covariance in the stationary regime (in the absence of external shocks $X(t)$) between $y_i(t)$ and $y_j(t)$ is denoted by $\sigma_{ij}(t)= \langle ( y_i(t) -\langle y_i \rangle ) ( y_j(t) - \langle y_j \rangle ) \rangle$, and $\sigma_{ij}(t \to \infty) \equiv \sigma_{ij}$.
By the central limit theorem, the stationary solution for this process is immediately given by
\begin{equation}
f_0(y) = \frac{1}{\sqrt{(2\pi)^N |\sigma|}} \exp \left(-\frac{1}{2} (\sigma^{-1})_{ij} y_i y_j  \right) \quad.
\end{equation}
The perturbation is of the form $L_X(y,t) = L_X(y_i) X_i(t)$.
We can identify the time evolution operator, $L_X(y_i)$, and the generalized potential, $\phi$, as
\begin{equation}
L_X(y_i) = \frac{\partial}{\partial y_i} \ ; \ \ \frac{\partial \phi}{\partial y_i} = (\sigma^{-1})_{ij} y_j \quad.
\end{equation}
The response of component $y_k$ to an external force acting on component $i$ is then
\begin{equation}
R_{y_k,X_i}(t) = (\sigma^{-1})_{ij} \langle y_k(t) y_j(0) \rangle \cdot X_i(t) \quad.
\end{equation}

We derived these results making the following three assumptions, namely that our system is (i) linear and (ii) time invariant and that (iii) the external field only depends on time.
In addition, we only considered correction terms of first order in Eq. \ref{approxim} and therefore neglect potential non-linear higher order effects.
The applicability of the central limit theorem to the stationary solution is another consequence of these assumptions.
Note that the fact that many economic time series follow fat-tailed or power law distributions is not at variance with the property that the stationay solution is given by a multivariate normal distribution.
After all, there is no reason to assume that the stationary solution $f_0(y)$ can at any point in time be actually observed in the data due to the ceaseless impacts of direct and indirect shocks; $f_0(y)$ is a computational crutch.

\subsection{Note S6: On the difference between the susceptibility matrix and the Leontief inverse matrix}

In this note we clarify the difference between the susceptibilty matrix $\rho$ and the Leontief inverse matrix.
In particular, for a step demand shock in LRT we find a non-equilibrium stationary state given by the linear relation $\langle \Delta \mathbf{Y}  \rangle_X = \mathbf{\rho X}$,
whereas in the ``standard'' Leontief IO economy we would expect a perturbed equilibrium state given by $\Delta \mathbf{Y} = (\mathbb{I}-\mathbf{A})^{-1} \mathbf{X}$.
Wherein lies the fundamental difference between these two expressions?

To obtain the LRT solution for the output change, we describe the Leontief IO model in terms of a stochastic differential equation, see Eq. (5) in the main text,
\begin{equation}
\dot \mathbf{Y} = (\mathbf{A}-\mathbb I)\mathbf{Y} + \mathbf{D} + \mathbf{X}(t) + \mathbf{F}(t) \quad.
\label{SLIOM2}
\end{equation}
The formal and stationary solution of this equation in the absence of a shock ($\mathbf{X}(t)=0$!) is given in the Supporting Note S5, Eq.(32); it is a probability density function (where we changed variables to have a homogeneous equation, $\mathbf{Y} \to \mathbf{y} = \mathbf{Y} - (\mathbb I - \mathbf{A})^{-1}\mathbf{D}$),
\begin{equation}
\mathbf{f_0}(\mathbf{y}) = \frac{1}{\sqrt{(2\pi)^N |\mathbf{\sigma}|}} \exp \left(-\frac{1}{2} (\mathbf{y} \mathbf{\sigma}^{-1}) \mathbf{y}  \right) \quad.
\label{f_02}
\end{equation}
The LRT prediction is an expectation value computed using this probability density, which is a multivariate normal distribution centred on the solution of the unperturbed Leontief IO model, $(\mathbb I - \mathbf{A})^{-1}\mathbf{D}$, when expressed in terms of the variable $\mathbf{Y}$.
For a step demand shock we get Eq. (9) as stationary solution ($t \to \infty$),
\begin{equation}
\langle \Delta Y_k  \rangle_X = \rho_{ki} X_i, \ \mathrm{with} \ \rho_{ki} = \int_0^{\infty} (\sigma^{-1})_{ij} \langle Y_k(\tau) Y_j(0) \rangle_0 d\tau \quad, 
\label{LinLaw2}
\end{equation}
where $\langle \cdot \rangle_0$ means that the expectation value is computed assuming $\mathbf{f_0}$ as underlying probability density function.
We emphasize again that Eq. (\ref{LinLaw2}) shows how a {\em non-equilibrium} expectation value ($\langle \Delta Y_k  \rangle_X$ for $\mathbf{X} \neq \mathbf{0})$ can be defined in terms of a known {\em equilibrium} expectation value, namely $\rho_{ki}$ in Eq. (\ref{LinLaw2}).

The computation of output changes in a perturbed Leontief IO model would proceed along a different route.
Let us again consider a step demand shock $\mathbf{X}(t) =\mathbf{X} \theta(t)$.
We are now interested in the {\em equilibrium} state of an economy under such a demand shock.
Therefore, we introduce the perturbed demand, $\mathbf{D_P}$ as $\mathbf{D_P} = \mathbf{D}+\mathbf{X}$ and ask for the stationary solution of the stochastic differential equation,
 \begin{equation}
\dot \mathbf{Y} = (\mathbf{A}-\mathbb I)\mathbf{Y} + \mathbf{D_P} + \mathbf{F}(t) \quad.
\label{SLIOM_PER}
\end{equation}
Formally, Eqs. (\ref{SLIOM2}) and  (\ref{SLIOM_PER}) are identical.
However, the perturbed equilibrium perspective assumes a different stationary solution, namely
\begin{equation}
\mathbf{f_P}(\mathbf{z}) = \frac{1}{\sqrt{(2\pi)^N |\mathbf{\sigma_P}|}} \exp \left(-\frac{1}{2}  (\mathbf{z} \mathbf{\sigma_P}^{-1}) \mathbf{z}  \right) \quad,
\label{f_P}
\end{equation}
where  $\mathbf{z} = \mathbf{Y} - (\mathbb I - \mathbf{A})^{-1}\mathbf{D_P} \neq  \mathbf{Y} - (\mathbb I - \mathbf{A})^{-1}\mathbf{D}$ and $(\sigma_P)_{ij}(t)= \langle ( z_i(t) -\langle z_i \rangle ) ( z_j(t) - \langle z_j \rangle ) \rangle$ with $\mathbf{\sigma_P}(t \to \infty) \equiv \mathbf{\sigma_P}$.
As could be expected, the stationary state of Eq.(\ref{SLIOM_PER}), i.e. the distribution of values of $\Delta \mathbf{Y}$ for $t \to \infty$, is now a multivariate normal distribution around the perturbed equilibrium state $(\mathbb I-\mathbf{A})^{-1}\mathbf{D_P}$.
It is clear that this solution coincides with the LRT solution for $\mathbf{D_P}=\mathbf{D}$, i.e. in the absence of a shock, $\mathbf{X}=0$.
Also, only for $\mathbf{X}=0$ would the correct expectation value to compute the susceptibility matrix and output change in LRT be given by the distribution $\mathbf{f_P}(\mathbf{z})$.

Within this note we have now encountered three different types of expectation value, namely (i) the equilibrium expectation value using measure $\mathbf{f_0}$,  $\langle \cdot \rangle_0$, (ii) the perturbed equilibrium expectation value given by $\mathbf{f_P}$,  call it $\langle \cdot \rangle_P$, and (iii) the nonequilibrium expectation value $\langle \cdot \rangle_X$.
These three expectation values coincide only in the absence of shocks, $\mathbf{X}=0$.
From a physical point of view, they describe three different types of system, namely a system relaxing to the state (i) $\mathbf{f_0}$, (ii) $\mathbf{f_P}$, or (iii) $\mathbf{f_0}$ while responding to an external driving force, $\mathbf{X}(t)$ (a physicist would say that the external force $\mathbf{X}$ ``does work on the system'').
In physics, the latter class of systems are closely related to ``dissipative structures'', i.e. systems in a steady non-equilibrium state driven by an exchange of energy and/or matter with the environment.

In brief, while both the LRT and perturbed IO approach start from the same stochastic differential equation, Eqs. (\ref{SLIOM2}) and  (\ref{SLIOM_PER}), they fundamentally differ in their definitions of expectation values.
In constrast to expected values in the Leontief IO model in the perturbed equilibrium approach, the LRT approach assumes that the stationary solution of the system does  {\em not} change after the application of an external shock.
It might also be instructive to compare predictions of the LRT model with those from the perturbed Leontief IO model, i.e. the prediction $\langle \Delta Y(t+\Delta t) \rangle_P = (\mathbb I-\mathbf{A})^{-1}\mathbf{\tilde X}$, where $\mathbf{\tilde X}$ is the implied shock from Eq. (\ref{ImpliedShock}).
We have to emphasize that such a comparison is problematic, as the use of the implied shock is only properly defined within the LRT framework, whereas the forecast using the perturbed IO model assumes a different type of dynamical system, as described above.
In this sense, these two models are incompatible and should not be treated on equal footing as we do it here.
Nevertheless, using the same evaluation strategy as we did for comparisons with time series models, we find that the LRT predictions significantly outperform predictions from the perturbed IO model ($p<10^{-90}$); see also the Supplementary Figure 10.

\begin{figure*}[tbp]
\begin{center}
 \includegraphics[width = 0.9\textwidth, keepaspectratio = true]{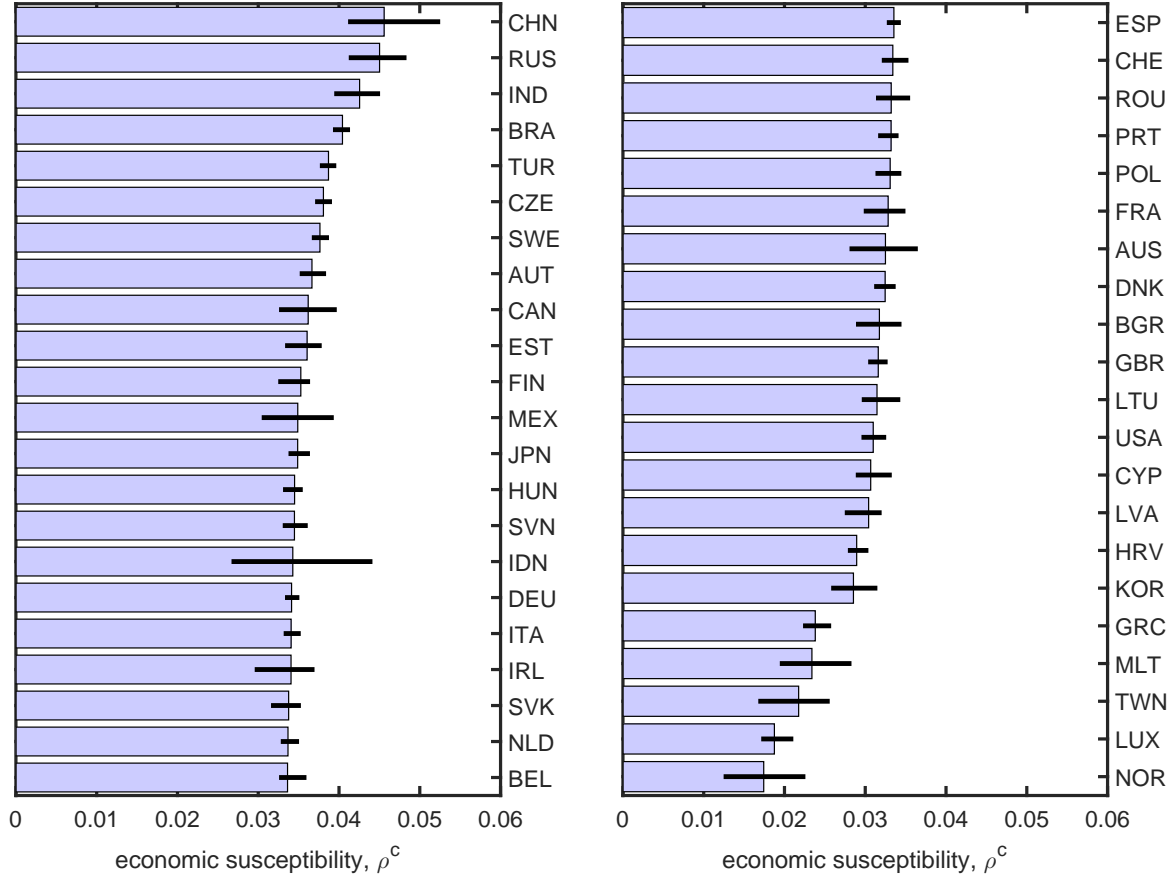}
\end{center}
 \caption{Supplementary Figure 1: Ranking of countries according to their susceptibility to economic shocks, calculated as the average over the susceptibility of each of the country's sectors. Error bars indicate the standard deviation of the sectoral susceptibilities.}
 \label{coun}
\end{figure*} 

\begin{figure*}[tbp]
\begin{center}
 \includegraphics[width = 0.9\textwidth, keepaspectratio = true]{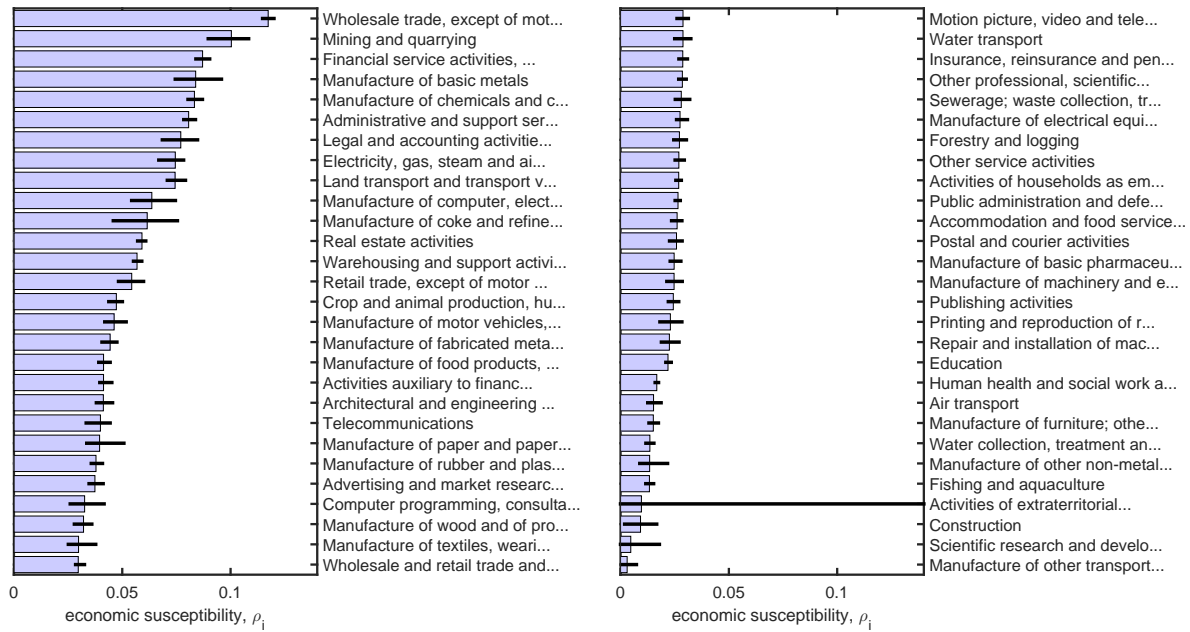}
\end{center}
 \caption{Supplementary Figure 2: Ranking of sectors according to their susceptibility to economic shocks, calculated as the average over the susceptibility of each of the corresponding sector in each of the countries.}
 \label{sec}
\end{figure*}

\begin{figure*}[tbp]
\begin{center}
 \includegraphics[width = 0.9\textwidth, keepaspectratio = true]{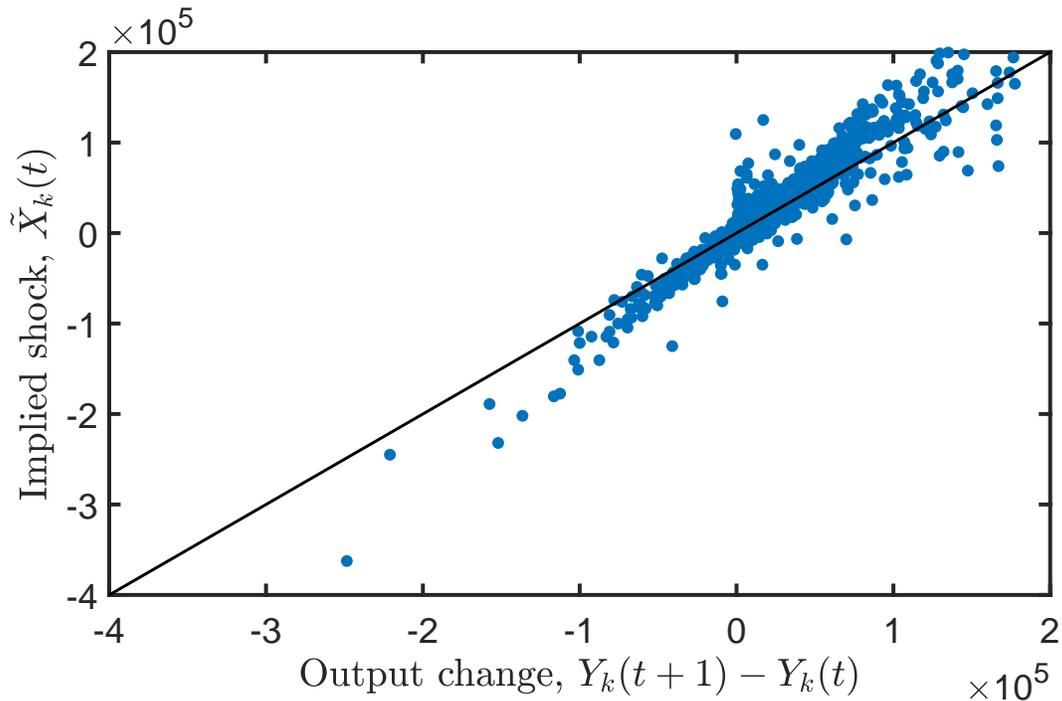}
\end{center}
 \caption{Supplementary Figure 3: Comparison of the observed output changes, $Y_k(t+1)-Y_k(t)$, and the implied step demand shock $\tilde X_k(t)$. Each sector from each country is shown for each year between 2003 and 2014 as a blue circle, the black solid line shows $\tilde X_k(t) = Y_k(t+1)-Y_k(t)$. There is a clear tendency that the implied shocks are larger in absolute value than the observed output changes, meaning that the response formalism typically attenuates the initial shock. There are, however, outliers that defy this general tendency. For instance, $t=2008$, the largest negative implied shock was observed for the manufacturing of coke sector in the US.}
 \label{impsh}
\end{figure*}

\begin{figure*}
\begin{center}
 \includegraphics[width = 0.9\textwidth, keepaspectratio = true]{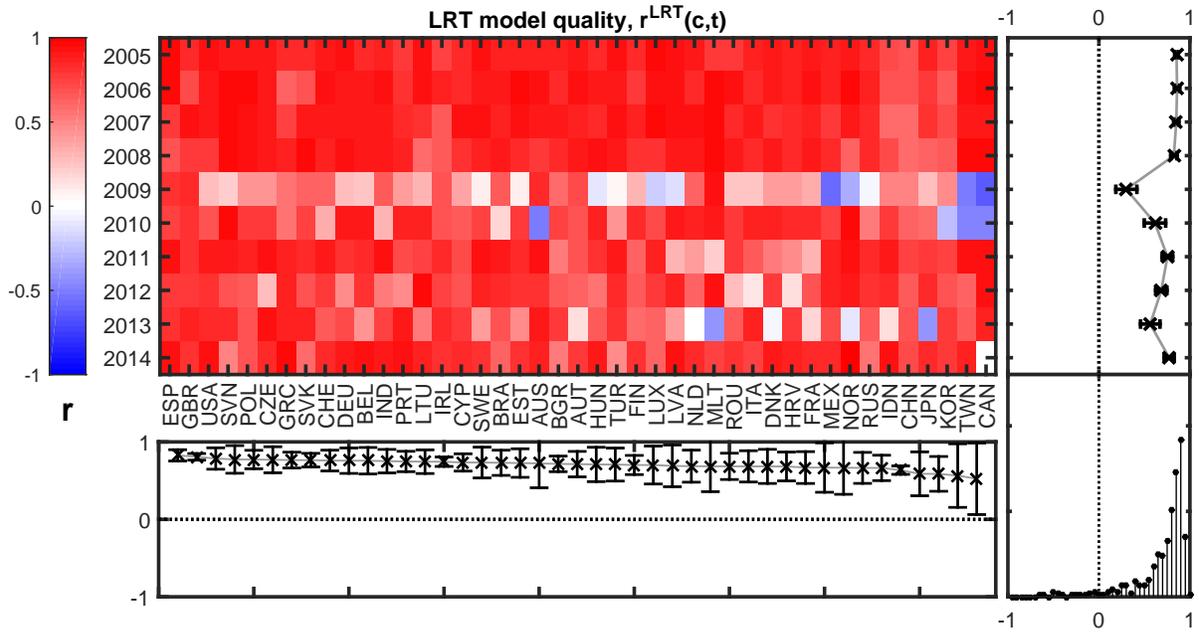}
\end{center}
 \caption{Supplementary Figure 4: Values of the correlation coefficient $r^{LRT}(c,t)$ between actual output changes and the predictions of the LRT model for each country $c$ and year $t$ (top left panel). 
 We show $r^{LRT}(c,t)$ averaged over each year (top right) and country (bottom left). Averages that are significantly different from zero are highlighted in bold and black.  
A histogram of $r^{LRT}(c,t)$ over all countries and years (bottom right) shows the corresponding distribution.
We find values of $r^{LRT}(c,t)$ close to the maximal value of $1$ for almost all countries and years except 2009.} 
 \label{rLRT}
\end{figure*} 

\begin{figure*}
\begin{center}
 \includegraphics[width = 0.9\textwidth, keepaspectratio = true]{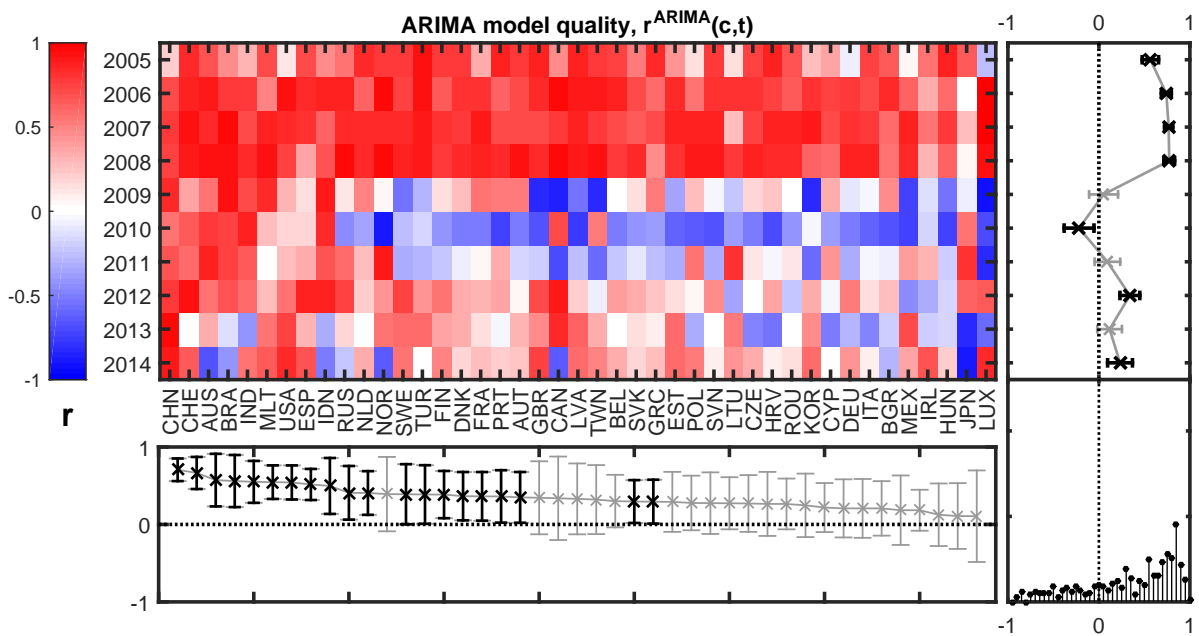}
\end{center}
 \caption{Supplementary Figure 5: Same as Supporting Fig. 5 for the correlation coefficient $r^{ARIMA}(c,t)$ between actual output changes and the predictions of the ARIMA model.
We find values of $r^{ARIMA}(c,t)$ close to the maximal value of $1$ up until 2008. In later years the average correlation coefficients fluctuate around zero.} 
 \label{rARIMA}
\end{figure*}

\begin{figure*}
\begin{center}
 \includegraphics[width = 0.7\textwidth, keepaspectratio = true]{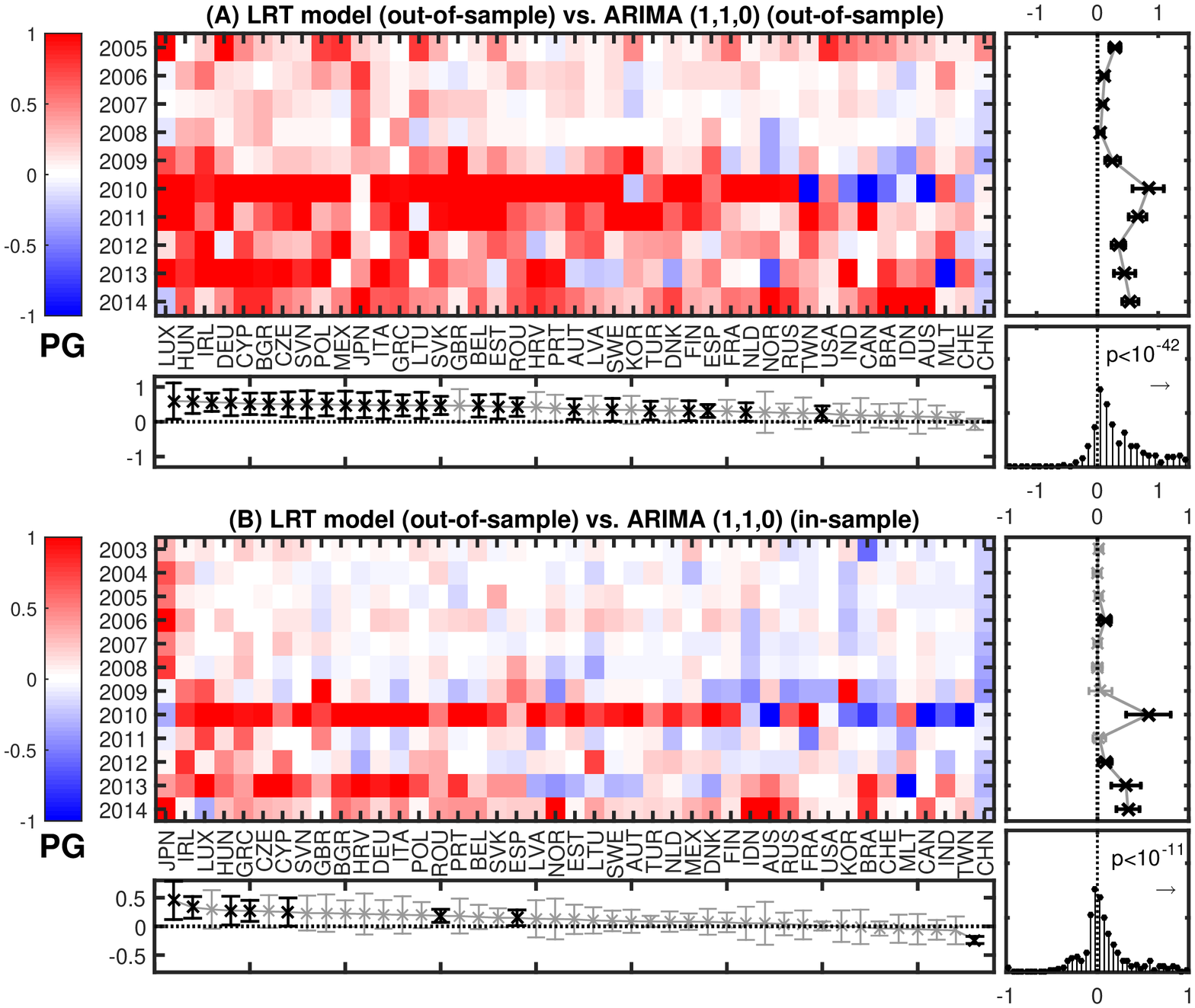}
\end{center}
 \caption{Supplementary Figure 6: Same as Fig. \ref{pg_pure} for comparison of the LRT versus an ARIMA(1,1,0) model.}
 \label{pure110}
\end{figure*} 

\begin{figure*}
\begin{center}
 \includegraphics[width = 0.7\textwidth, keepaspectratio = true]{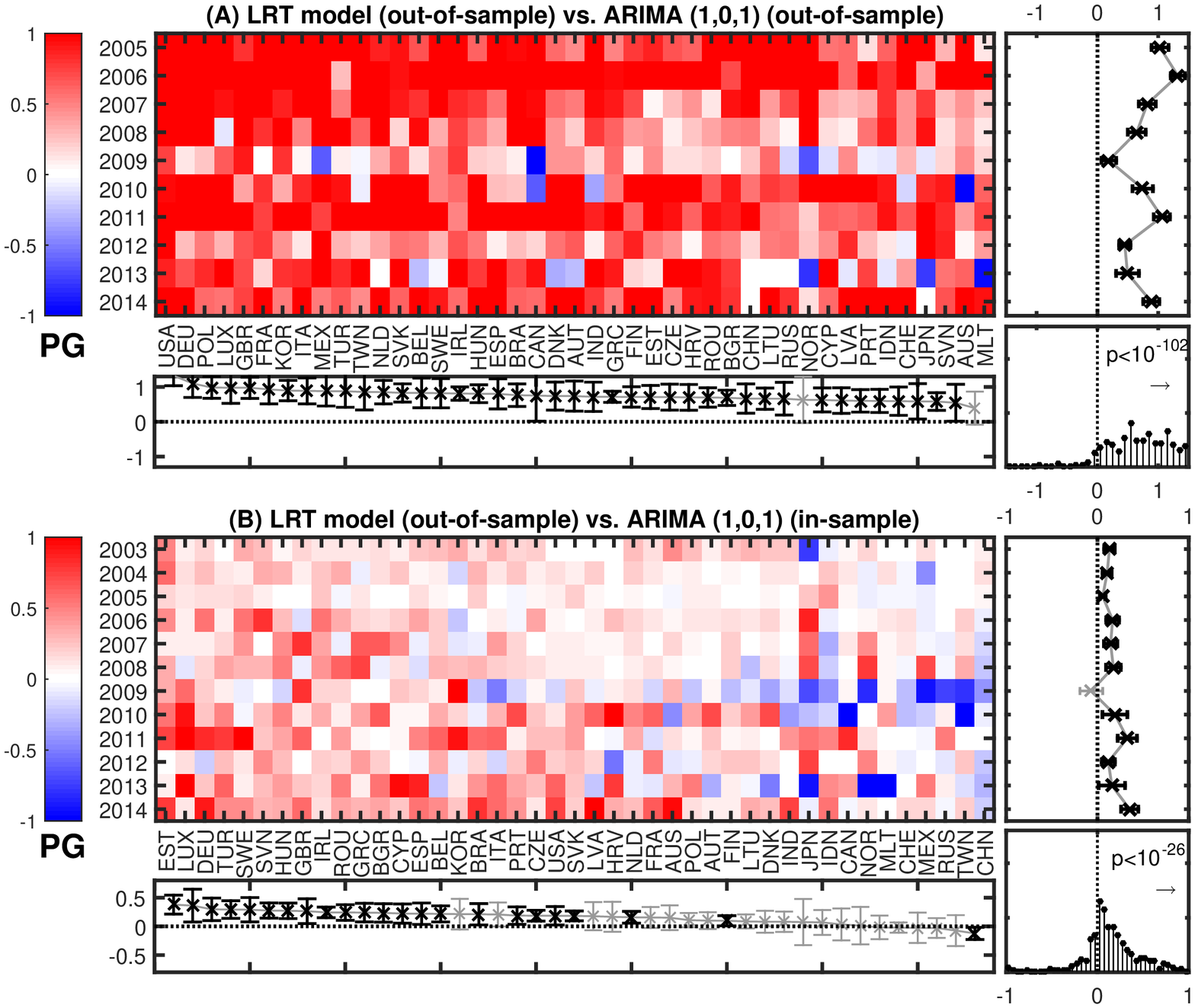}
\end{center}
 \caption{Supplementary Figure 7: Same as Fig. \ref{pg_pure} for comparison of the LRT versus an ARIMA(1,0,1) model. }
 \label{pure101}
\end{figure*} 

\begin{figure*}
\begin{center}
 \includegraphics[width = 0.7\textwidth, keepaspectratio = true]{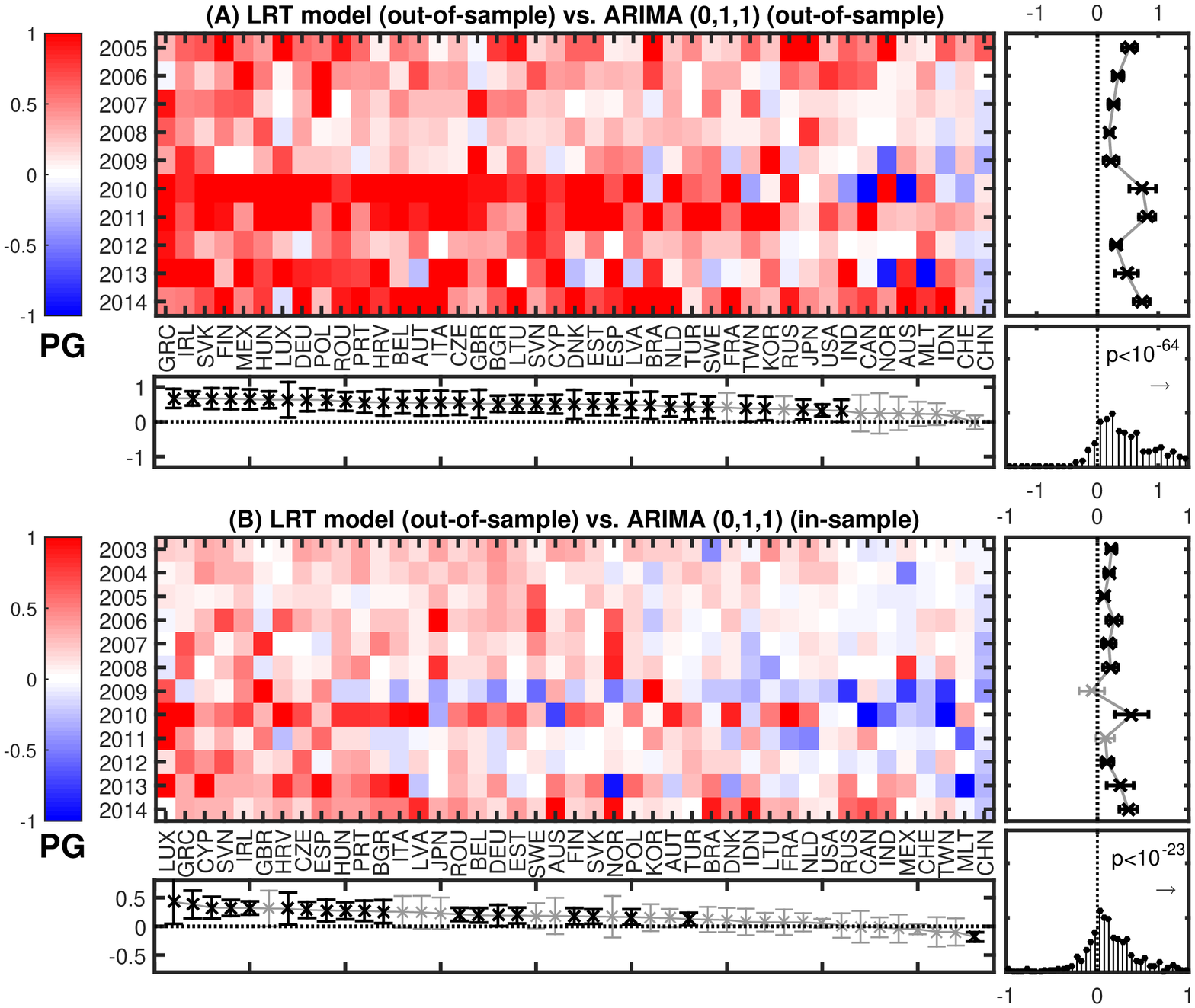}
\end{center}
 \caption{Supplementary Figure 8: Same as Fig. \ref{pg_pure} for comparison of the LRT versus an ARIMA(0,1,1) model.}
 \label{pure011}
\end{figure*}

\begin{figure*}
\begin{center}
 \includegraphics[width = 0.9\textwidth, keepaspectratio = true]{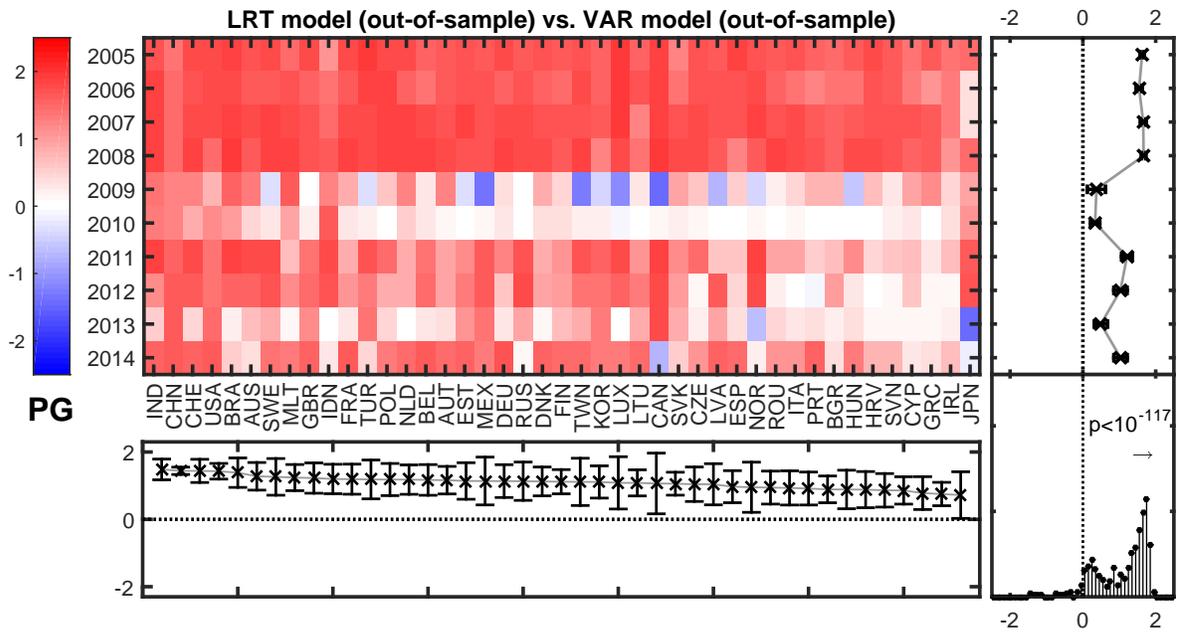}
\end{center}
 \caption{Supplementary Figure 9: Comparison of the predictions of the linear response model with a sectoral VAR model. 
 (A) We compare the LRT model for a shock between years $t$ and $t+1$ with VAR models that have been calibrated 
 using data from the year 2000. For each country and each year, we show the difference in Pearson's correlation coefficients between observed output changes and the different types of models, the  predictability gain, $PG$. We show $PG$ averaged over each year (panel to the right) and country (bottom panel). Here averages that are significantly different from zero  are highlighted in bold and black.  
A histogram of $PG$ over all countries and years (bottom right) shows the corresponding distribution.
 We find that in terms of predictability gain $PG$ the LRT model is vastly superior to the VAR model ($p<10^{-117}$).
The advantage is least pronounced (but still significant) in 2009 and 2010, whereas values for $PG$ in the other years are typically between one and two.} 
 \label{VARm}
\end{figure*} 

\begin{figure*}
\begin{center}
 \includegraphics[width = 0.9\textwidth, keepaspectratio = true]{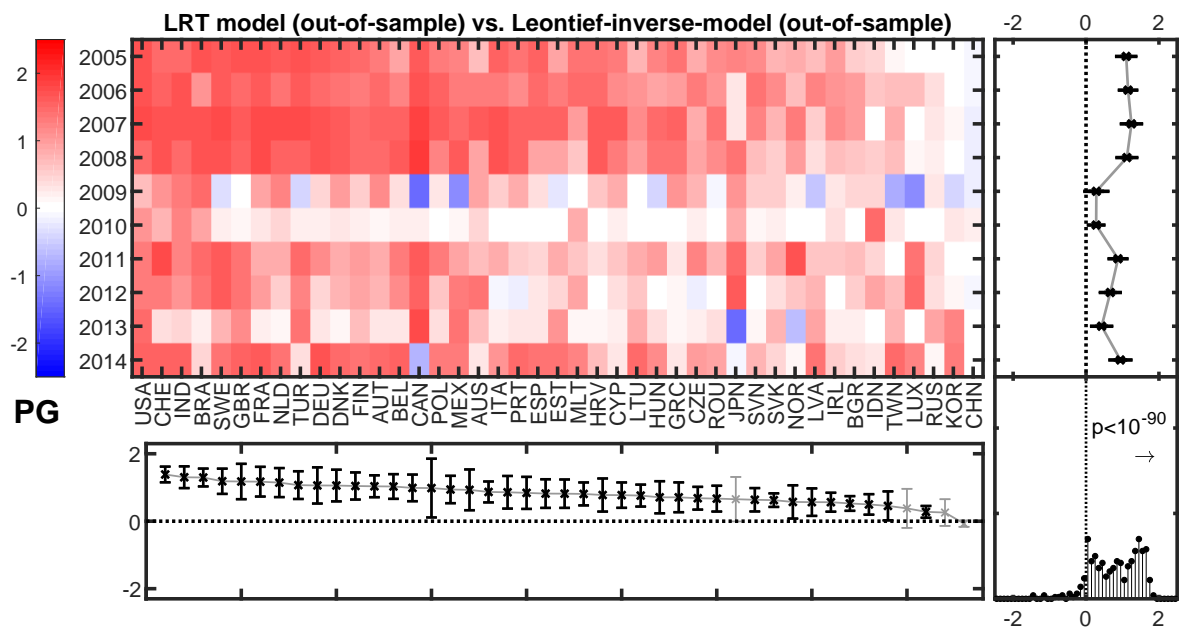}
\end{center}
 \caption{Supplementary Figure 10: A comparison of the LRT model versus predictions from the perturbed equilibrium IO model.}
 \label{pure011}
\end{figure*}

\begin{table*}[b]
\caption{Supplementary Table 1: Many phenomenological laws can be understood by linear response theory. Each of these laws is of the form $J=\rho X$ with $X$ being a perturbation that induces a flux $J$. The size of the response $J$ is proportional to the perturbation $X$. The linear dependence is described by transport coefficients $\rho$ which can be a scalar, a vector, or a tensor.}
\begin{tabular}{l | l | l | l}
phenomenological law & perturbation $X$ & induced flux $J$ & transport coefficient $\rho$ \\
\hline
$I=\sigma V$ & voltage $V$ & current $I$ & resistance $\sigma$ \\
$S_{xy} = \sum_{kl}\mu_{ijkl}E_{kl}$  & strain rate $E$ & viscuous stress $S$ & viscosity tensor $\mu$ \\
$B_i = \mu H_i$ & magnetic field $H$ & magnetic flux density $B$ & magnetic susceptibility $\mu$ \\
$E_i = \epsilon_{ij} D_i$ & electric field $E$ & electric displacement $D$ & electric susceptibility $\epsilon$ \\
$\Delta Y_k = \rho_{ki} X_i$ & demand shock $X$ & output change $\Delta Y$ & economic susceptibility $\rho$
\end{tabular}
\label{LRTLaws}
\end{table*}

\begingroup
\squeezetable
\begin{table*}[h]
\caption{Supplementary Table 2: List of sectors included in the WIOD. For each sector we give its ISIC code(s), full name, short name, the type it was assigned to in Fig. 1, and its susceptibility score with 95\% confidence interval.}
\begin{tabular}{p{1.1cm} | p{8cm} | p{3cm} | p{2.2cm} | l}
code & full name & short name & type & susceptibility, $\rho_i$ (95\% CI) \\
\hline
A01 & Crop and animal production, hunting and related service activities &	Agriculture &	Agriculture & 0.0382 (0.0364--0.0404)  \\
A02  & Forestry and logging &	Forestry &	Agriculture &  0.0290 (0.0270--0.0305) \\
A03 & Fishing and aquaculture &	Fishing &	Agriculture &  0.0183 (0.0177--0.0196) \\
B & Mining and quarrying &	Mining &	Mining &  0.0568 (0.0505--0.0620) \\
C10-C12 & Manufacture of food products, beverages and tobacco products &	Food &	Manufacturing & 0.0355 (0.0336--0.0369) \\
C13-C15 & Manufacture of textiles, wearing apparel and leather products &	Textiles &	Manufacturing &  0.0226 (0.0213--0.0240) \\
C16 & Manufacture of wood and of products of wood and cork, except furniture; manufacture of articles of straw and plaiting materials &	Wood &	Manufacturing & 0.0270 (0.0259--0.0284) \\
C17 & Manufacture of paper and paper products &	Paper &	Manufacturing & 0.0295 (0.0267--0.0338) \\
C18 & Printing and reproduction of recorded media &	Printing &	Manufacturing & 0.0284 (0.0248--0.0319) \\
C19 & Manufacture of coke and refined petroleum products &	Coke &	Manufacturing & 0.0473 (0.0409--0.0519) \\
C20 & Manufacture of chemicals and chemical products &	Chemicals &	Manufacturing & 0.0450 (0.0428--0.0497) \\
C21 & Manufacture of basic pharmaceutical products and preparations &	Pharmaceuticals &	Manufacturing & 0.0191 (0.0184--0.0205) \\
C22 & Manufacture of rubber and plastic products &	Rubber &	Manufacturing & 0.0285 (0.0261--0.0314) \\
C23 & Manufacture of other non-metallic mineral products &	Mineral products &	Manufacturing & 0.0248 (0.0200--0.0309) \\
C24 & Manufacture of basic metals &	Metals &	Manufacturing & 0.0576 (0.0505--0.0643) \\
C25 & Manufacture of fabricated metal products, exc. machinery \& equipment &	Metal products &	Manufacturing & 0.0387 (0.0361--0.0423) \\
C26 & Manufacture of computer, electronic and optical products &	Computer &	Manufacturing & 0.0292 (0.0275--0.0311) \\
C27 & Manufacture of electrical equipment &	Electricals &	Manufacturing & 0.0207 (0.0185--0.0241) \\
C28 & Manufacture of machinery and equipment n.e.c. &	Machinery &	Manufacturing & 0.0202 (0.181--0.0228) \\
C29 & Manufacture of motor vehicles, trailers and semi-trailers &	Motor vehicles &	Manufacturing & 0.0246 (0.0220--0.0267) \\
C30 & Manufacture of other transport equipment &	Transport equ. &	Manufacturing & 0.0127 (0.0112--0.0143) \\
C31-32 & Manufacture of furniture; other manufacturing &	Furniture &	Manufacturing & 0.0193 (0.0178--0.0208) \\
C33 & Repair and installation of machinery and equipment &	Repair &	Manufacturing & 0.0324 (0.0287--0.0348) \\
D35 & Electricity, gas, steam and air conditioning supply &	Electricity &	Electricity \& Water & 0.0699 (0.0638--0.0736) \\
E36 & Water collection, treatment and supply &	Water &	Electricity \& Water  & 0.0183 (0.0175--0.0195) \\
E37-E39 & Sewerage; waste collection, treatment and disposal activities; materials recovery; remediation activities and other waste management services &	Waste &	Electricity \& Water & 0.0275 (0.0256--0.0305) \\
F & Construction &	Construction &	Construction & 0.0246 (0.0138--0.0305) \\
G45 & Wholesale and retail trade and repair of motor vehicles and motorcycles &	Car trade &	Trade & 0.0322 (0-0307--0.347) \\
G46 & Wholesale trade, except of motor vehicles and motorcycles &	Wholesale trade &	Trade & 0.0866 (0.0808--0.0916) \\
G47 & Retail trade, except of motor vehicles and motorcycles &	Retail trade &	Trade & 0.0509 (0.0442--0.0553) \\
H49 & Land transport and transport via pipelines &	Land transport &	Transport & 0.0581 (0.0546--0.0620) \\
H50 & Water transport &	Water transport &	Transport & 0.0178 (0.0161--0.0199) \\
H51 & Air transport &	Air transport &	Transport & 0.0162 (0.0141-0.0180) \\
H52 & Warehousing and support activities for transportation &	Warehousing &	Transport & 0.0429 (0.0415--0.0445) \\
H53 & Postal and courier activities &	Post &	Transport & 0.0279 (0.0260--0.0296) \\
I & Accommodation and food service activities &	Accommodation &	Accommodation & 0.0270 (0.0256--0.0284) \\
J58 & Publishing activities &	Publishing &	Inform. \& Comm. & 0.0265 (0.0253--0.0277) \\
J59-J60 & Motion picture, video and television programme production, sound recording and music publishing activities; &	Entertainment &	Inform. \& Comm. & 0.0335 (0.0316--0.0352) \\
J61 & Telecommunications & 	Telecommunication &	Inform. \& Comm. & 0.0398 (0.0338--0.0434) \\
J62-J63 & Computer programming, consultancy and related activities &	Computer programming &	Inform. \& Comm. & 0.0298 (0.0277--0.0324) \\
K64 & Financial service activities, exc. insurance and pension funding &	Financial services &	Finance & 0.0656 (0.0630--0.0674) \\
K65 & Insurance, reinsurance \& pension funding, exc. compulsory social security &	Insurance &	Finance & 0.0266 (0.0253--0.0275) \\
K66 & Activities auxiliary to financial services and insurance activities &	Auxiliary financial serv. &	Finance & 0.0379 (0.0366--0.0391) \\
L68 & Real estate activities &	Real estate &	Other & 0.0579 (0.0563--0.0592) \\
M69-M70 & Legal and accounting activities; activities of head offices; management consultancy activities &	Legal activities &	Other & 0.0634 (0.0589--0.0664) \\
M71 & Architectural and engineering activities; technical testing and analysis &	Architecture &	Other & 0.0390 (0.0373--0.0409) \\
M72 & Scientific research and development &	Research &	Research & 0.0148 (0.0024--0.0222) \\
M73 & Advertising and market research &	Advertising &	Research & 0.0375 (0.0344--0.0398) \\
M74-M75 & Other professional, scientific and technical activities; veterinary activities &	Other technical activities &	Research & 0.0311 (0.0299--0.0321) \\
N & Administrative and support service activities & 	Administration &	Administration & 0.0707 (0.0686--0.0727) \\
O84 & Public administration and defence; compulsory social security &	Public administration &	Administration & 0.0264 (0.0256--0.0272) \\
P85 & Education &	Education	 & Other & 0.0235 (0.0226--0.0246) \\
Q & Human health and social work activities &	Health &	Other & 0.0195 (0.0187--0.0206) \\
R-S & Other service activities &	Other services &	Other  & 0.0291 (0.0280--0.0301) \\
T & Activities of households as employers; &	Household activities &	Other & 0.0272 (0.0259--0.0283) \\
U & Activities of extraterritorial organizations and bodies &	Extraterrestrial org. &	Other & -0.0396 (-0.2838--0.3025) \\
\end{tabular}
\label{SectorTable}
\end{table*}
\endgroup

\end{document}